\documentclass[useAMS,usenatbib]{mn2e}
\usepackage{amssymb}
\usepackage{amsmath}
\usepackage{graphicx}
\usepackage{color}
\usepackage{natbib}
\usepackage[T1]{fontenc} 
\usepackage{aecompl}

\newcommand{\dd}{\mathrm{d}}

\newcommand{\mbh}{\ensuremath{M_\bullet}\,}
\newcommand{\er}{\ensuremath{\lambda\,}}
\newcommand{\mbhbu}{\ensuremath{M_\bullet - M_\mathrm{Bulge}\,}}
\newcommand{\mbhsig}{\ensuremath{M_\bullet - \sigma_\ast\,}}

\newcommand{\apjs}{ApJS}
\newcommand{\apj}{ApJ}
\newcommand{\mnras}{MNRAS}
\newcommand{\nat}{Nature}
\newcommand{\aj}{AJ}
\newcommand{\aap}{A\&A}
\newcommand{\apjl}{ApJ}

\title[Infering the intrinsic BH-bulge relation]{Accounting for selection effects in the BH-bulge relations:  no evidence for cosmological evolution}
\author[A. Schulze and L. Wisotzki]{Andreas Schulze$^{1,2}$\thanks{E-mail:
andreas.schulze@ipmu.jp}  and Lutz Wisotzki$^{3}$\\
$^{1}$ Kavli Institute for Astronomy and Astrophysics, Peking University, 100871 Beijing, China \\
$^{2}$ Kavli Institute for the Physics and Mathematics of the Universe (Kavli IPMU, WPI), Todai Institutes for Advanced Study,\\ The University of Tokyo, Kashiwa 277-8583, Japan\\
$^{3}$ Leibniz-Institut f\"ur Astrophysik Potsdam (AIP), An der Sternwarte 16, 14482 Potsdam, Germany}
\begin{document}

\date{Accepted 2013 December 17.  Received 2013 December 17; in original form 2013 November 20 }

\pagerange{\pageref{firstpage}--\pageref{lastpage}} \pubyear{2013}

\maketitle

\label{firstpage}

\begin{abstract}
The redshift evolution of the black hole - bulge relations is an essential observational constraint for models of black hole - galaxy coevolution. In addition to the observational challenges for these studies, conclusions are complicated by the influence of selection effects. 
We demonstrate that there is presently no statistical significant evidence for cosmological evolution in the  $M_\bullet$-bulge relations, once these selection effects are taken into account and corrected for.
We present a fitting method, based on the bivariate distribution of black hole mass and galaxy property, that accounts for the selection function in the fitting and is therefore able to recover the intrinsic black hole - bulge relation unbiased. While prior knowledge is restricted to a minimum, we at least require knowledge of either the sample selection function and the mass dependence of the active fraction, or the spheroid distribution function and the intrinsic scatter in the black hole - bulge relation.
We employed our fitting routine to existing studies of the $M_\bullet-$bulge relation at $z\sim1.5$ and $z\sim6$, using our current best knowledge of the distribution functions. There is no statistical significant evidence for positive evolution in the  $M_\bullet-M_\ast$ ratio out to $z\sim2$.
At $z\sim6$ the current constraints are less strong, but we demonstrate that the large observed \textit{apparent} offset from the local $M_\bullet-$bulge relation at $z\sim6$ is fully consistent with no \textit{intrinsic} offset. The method outlined here provides a tool to obtain more reliable constraints on black hole - galaxy co-evolution in the future.
\end{abstract}

\section{Introduction}
The tight correlation between the mass of a supermassive black hole and the properties of its host galaxies' spheroid component is now well established in the local Universe \citep[e.g.][]{Magorrian:1998,Ferrarese:2000,Gebhardt:2000,Tremaine:2002,Marconi:2003, Haering:2004}, indicating a coeval growth history of galaxies and their central black holes. However, the detailed role of this correlation for our understanding of galaxy evolution and black hole growth is still an open question. The standard theoretical scenario explains this relation by joint triggering of star formation and black hole activity via major mergers, while the black hole growth is self-regulated via active galactic nuclei (AGN) feedback that shuts off star formation and quenches the accretion on to the black hole \citep[e.g.][]{Silk:1998,Kauffmann:2000,DiMatteo:2005,Sijacki:2007,Somerville:2008}. Alternatively, it may be possible  to reproduce the observed black hole - galaxy correlations without the need for self-regulation via AGN feedback, either as a natural consequence of a common merger history \citep{Peng:2007,Jahnke:2010}, or in a black hole growth scenario via gravitational torques \citep{Angles:2013}.

An observational constraint to distinguish between these alternative scenarios and to constrain  individual theoretical models \citep[e.g.][]{Croton:2006,Robertson:2006,Lamastra:2010,Booth:2010,Dubois:2011} is provided by the redshift evolution of the $M_\bullet-$bulge relations.
However, direct dynamical black hole mass measurements are not feasible at higher redshifts. While integrated constraints can be obtained from the study of the black hole mass function (BHMF) and the AGN luminosity function \citep{Merloni:2004, Hopkins:2006, Zhang:2012}, the only route for observational studies on individual objects is to use broad line AGN samples. 

\begin{figure}
\centering 
\resizebox{\hsize}{!}{\includegraphics[clip]{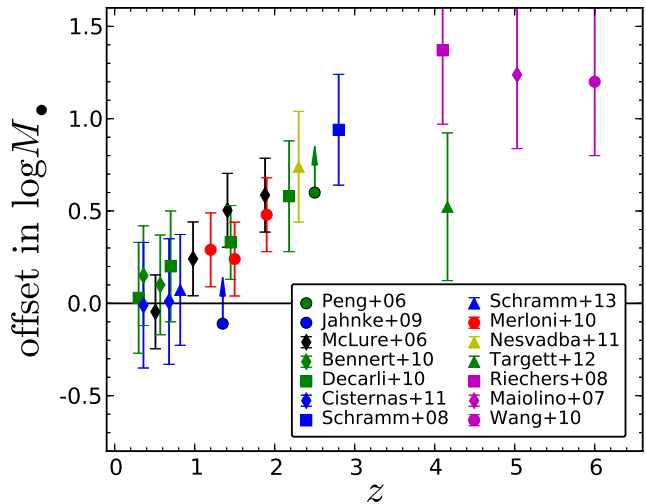}}
\caption{Compilation of results obtained from previous studies on the \textit{apparent} offset in black hole mass at a given galaxy mass from the local relation from \citet{Haering:2004}. Overall, an \textit{apparent} trend of an increasing $M_\bullet/M_\ast$ ratio is suggested. However, sample selection effects need to be considered to reconstruct the \textit{intrinsic} offset.}
\label{fig:bhmstevo}
\end{figure}

The underlying assumption for this approach is that broad line AGN obey the same relationship as non-active galaxies \citep[e.g.][]{Gebhardt:2000b,Woo:2013}.
For broad line AGN the black hole mass can be estimated employing the 'virial method' \citep[e.g.][]{McLure:2002,Vestergaard:2006}. The main challenge for these studies is to determine the properties of the AGN host galaxy, targeting either the velocity dispersion \citep{Shields:2003,Salviander:2007,Woo:2008,Canalizo:2012,Hiner:2012,Salviander:2013}, the galaxy luminosity \citep{Peng:2006b,Decarli:2010,Bennert:2010,Targett:2012} or the stellar mass \citep{McLure:2006,Schramm:2008,Jahnke:2009,Merloni:2010,Nesvadba:2011,Cisternas:2011}. While early  studies suggested a clear trend of positive evolution in the \mbh-bulge relation, they preferentially focused on luminous quasars and their hosts, sampling from the bright end of the AGN luminosity function. More recent studies, using fainter AGN, tend to find mild or no evolution \citep{Jahnke:2009,Merloni:2010,Cisternas:2011,Salviander:2013,Schramm:2013}. A compilation of current literature results is shown in Fig.~\ref{fig:bhmstevo}.

Indeed, sample selection is a particular important issue for these studies, as selection effects are almost inevitable \citep{Lauer:2007,Shen:2010,Schulze:2011b,Volonteri:2011,Portinari :2012}.  
In particular, \citet{Lauer:2007} argued that intrinsic scatter in the \mbh-bulge relation and a steep exponential cutoff in the galaxy distribution function will lead to biased apparent relations. This bias is exacerbated by the effects of a bright flux limit in the employed AGN samples.
In \citet[][hereafter SW11]{Schulze:2011b} we presented a common framework, based on the bivariate distribution function of black hole mass and spheroid property, to investigate and model the luminosity bias and other selection biases on the \mbh-bulge relations. We additionally discussed an active fraction bias, which is introduced if only \textit{active} black holes are selected and if the probability to be in an active stage depends on black hole mass. It is possible to account for these effects if (1) the selection function of the respective sample is well known, and (2)  the underlying distribution functions, such as the galaxy distribution function, the active BHMF and the Eddington ratio distribution function (ERDF), are known. However, in particular at high redshift this is currently not the case.

Therefore, we are facing the problem that even if we are able to obtain reasonably large AGN samples with individual black hole mass estimates and host galaxy property measurements, the interpretation of such results in terms of evolution or non-evolution is challenging due to the sample selection effects. In this paper we present a practical solution to this dilemma. 
We build on the framework presented in SW11 and give a quantitative assessment of, and a correction procedure for sample selection effects on observations of the \mbh-bulge relations. We outline a maximum likelihood fitting approach that directly incorporates information on the selection function into the fitting. We also restrict prior knowledge of the underlying distribution functions to a minimum. The method is able to reconstruct the true unbiased \mbh-bulge relation for an observed sample and thus largely overcomes the limitations imposed by the sample selection.

The paper is organized as follows: In section~\ref{sec:method} we present the general maximum likelihood fitting method and its application to the \mbh-bulge relation for AGN. In section~\ref{sec:predict} we verify the robustness and uncertainties of the method via Monte Carlo simulations.
Section~\ref{sec:application} illustrates the method by its application to representative previous studies and discusses its ramifications for the evolution of the \mbh-bulge relation.  We  conclude in section~\ref{sec:conclusions}. 

\section{The method}  \label{sec:method}
\subsection{Maximum likelihood fit in consideration of a selection function}
While the main goal of the method presented here is to study the  $M_\bullet-$bulge relation, it is also applicable to more general cases. Identifying and characterizing correlations between two observables is a fundamental aspect of astrophysical research. Given a joint distribution of two parameters with non-zero intrinsic scatter, this correlation will often be affected by the specific sample selection. 
We therefore first outline the general concept of the approach, before applying it to the $M_\bullet-$bulge relation.

Let us assume that there exists a correlation between two observational parameters $x$ and $y$ for a class of objects. The space density of these objects is given by a bivariate distribution function $\Psi(x,y)$. Marginalizing $ \Psi(x,y)$ over $x$ gives the distribution function of the variable $y$, and vice versa. In the case of the $M_\bullet-$bulge relation $x$ is for example the stellar velocity dispersion $\log \sigma_\ast$ and $y$ is the black hole mass $\log M_\bullet$. Thus marginalizing their bivariate distribution function over $\log \sigma_\ast$ gives the BHMF, while marginalizing over $\log M_\bullet$ will provide the galaxy stellar velocity dispersion function (SVDF).

We here assume the following parameterization for the bivariate distribution
\begin{equation}
 \Psi(x,y)=g(y \, |\, x)\, \Phi_x(x) \ ,  \label{eq:psi}
\end{equation} 
where $\Phi_x(x)$ is the distribution function in $x$, e.g. the galaxy SVDF, and $g(y \, |\, x)$ gives the probability of finding the value $y$ given $x$. This is the intrinsic correlation between $x$ and $y$.
We  here specifically assume a linear relation $y= \alpha + \beta x$ with log-normal intrinsic scatter $\sigma$, i.e.
\begin{equation}
 g(y \, |\, x) = \frac{1}{\sqrt{2 \pi} \sigma} \exp \left\lbrace - \frac{(y-\alpha-\beta x)^2}{2 \sigma^2} \right\rbrace  \ . \label{eq:gms}
\end{equation} 
The bivariate distribution function $\Psi(x,y)$ describes the intrinsic joint distribution function of $x$ and $y$ in the universe. In practice, this will not be equal to the observed bivariate distribution for a sample, since this sample is almost always affected by implicit or explicit selection effects. We incorporate these selection criteria into a selection function $\Omega(x,y,\theta)$, where $\theta$ refers to the set of additional parameters present as selection criteria, such as redshift or luminosity.
The multivariate distribution $\Psi(x,y,\theta)$ for an observed sample is then given by
\begin{equation}
  \Psi_\mathrm{o}(x,y,\theta)= \Omega(x,y,\mathbf{\theta})\, \Psi(x,y)\, p_\theta(\theta) \ , \label{eq:psio}
\end{equation} 
where $p_\theta(\theta)$ is a set of normalized distribution functions of the parameters $\theta$. The respective bivariate distribution $\Psi_\mathrm{o}(x,y)$ is given by integration over $\theta$. In general  $\Psi_\mathrm{o}(x,y) \neq \Psi(x,y)$. This means the correlation between $x$ and $y$, e.g. the $M_\bullet-\sigma_\ast$ relation, fitted directly to the observations is not equal to the intrinsic correlation. Ignoring this inequality will lead to a bias (see SW11).

In practice, we are mainly interested in the intrinsic correlation between $x$ and $y$, i.e. $g(y \, |\, x)$ and less in the full joint distribution function. The information of this intrinsic correlation is contained in the conditional probability function, but in general is not equal to it.
There are two options to define the observed  correlation between $x$ and $y$. First, the relation of $y$ at a given $x$, and second the relation of $x$ at a given $y$. These are  given by the conditional probabilities 
\begin{equation}
 p(y \, |\, x, \theta) = \frac{\Psi_\mathrm{o}(x,y,\theta)}{\int \Psi_\mathrm{o}(x,y,\theta)\, \dd y} \label{eq:cond_mus}
\end{equation} 
\begin{equation}
 p(x \, |\, y, \theta) = \frac{\Psi_\mathrm{o}(x,y,\theta)}{\int \Psi_\mathrm{o}(x,y,\theta)\, \dd x} \ , \label{eq:cond_smu} 
\end{equation} 
where $p(y \, |\, x, \theta)$ gives the probability of finding the property $y$ (e.g. black hole mass) in a galaxy with properties $x$ (e.g. $\sigma_\ast$) and $\theta$, while $p(x \, |\, y, \theta)$ is the probability of finding a galaxy with $x$ for given $y$ and $\theta$. We will refer to these as forward and inverse probabilities, respectively.

Let us assume the special case there are no selection effects on $y$.  When we use the bivariate distribution given by Equation~(\ref{eq:psi}) the forward conditional probability is given by $p_\mathrm{int}(y \, |\, x)=g(y \, |\, x)$, i.e. the intrinsic underlying relation is directly recovered in the conditional probability. For the inverse conditional probability it follows in the absence of any selection effects
\begin{equation}
 p_\mathrm{int}(x \, |\, y) = \frac{g(y \, |\, x)\, \Phi_x(x) }{\int g(y \, |\, x)\, \Phi_x(x) \dd x} \ .\label{eq:cond_smu_int} 
\end{equation} 
This is already a deviation from the intrinsic relation $g(y \, |\, x)$, even without applying any selection criteria to the sample. In the context of the $M_\bullet-$bulge relation the stellar mass function exponentially decreases at the high mass end, which together with the intrinsic scatter leads to a deviation of  $p_\mathrm{int}(M_\ast \, |\, M_\bullet)$ at the high mass end \citep[][SW11]{Lauer:2007}. This can already qualitatively explain the steeper slope found in the $M_\bullet-\sigma_\ast$ relation for galaxies with dynamical black hole mass measurements when fitting the inverse $M_\bullet-\sigma_\ast$ relation $p(\sigma_\ast \, |\, M_\bullet)$ instead of the relation $p(M_\bullet\, |\, \sigma_\ast)$ \citep{Graham:2011,Park:2012}.

In practice selection effects are present and can produce a bias, i.e. $p(y \, |\, x) \neq g(y \, |\, x)$. However, if we know the selection function of our sample, we can incorporate this knowledge directly into the fitting of the conditional probability to recover the intrinsic relation $g(y \, |\, x)$. The maximum likelihood technique provides a conceptually simple way to achieve this. This method aims at minimizing the likelihood function $S=-2\ln \mathcal{L}$, with $\mathcal{L}= \prod_i l_i$ being the product of the likelihoods for the individual measurements. 

Both conditional probabilities, $p(y \, |\, x) $ and $p(x \, |\, y)$, can serve as likelihoods, i.e. we either minimize
\begin{equation}
 S_x=-2 \sum_{i=1}^{N} \ln p(y_i \, |\, x_i, \theta_i) \ , \ \mathrm{or} \label{eq:Sx} 
\end{equation} 
\begin{equation}
 S_y=-2 \sum_{i=1}^{N} \ln p(x_i \, |\, y_i, \theta_i) \ , \label{eq:Sy} 
\end{equation} 
to obtain the best-fitting parameters of $g(y \, |\, x)$. We  here refer to the minimization of $S_x$  as forward ML regression and of $S_y$ as inverse ML regression. We emphasize that these maximum likelihood fits go beyond a simple linear regression, since we are not only fitting a model to the data, but incorporate additional information about the data set and the underlying distributions. Thereby, the used conditional probability function can have an arbitrarily complex form. We will refer to the more standard approach of fitting the conditional probability $p(\mu | s)$ without including the effects of sample selection as 'simple regression'.

Which of the two maximum likelihood fit options (forward/inverse) is the preferred choice depends on the specific selection effects, and on the underlying distributions. For example, when there are no selection effects then forward regression is preferable, since the fitted conditional probability is equal to the intrinsic relation $g(y \, |\, x)$. In this case the maximum likelihood fit is equal to a simple regression approach, used in other studies to fit the $M_\bullet-\sigma_\ast$ relation for quiescent galaxies \citep[e.g.][]{Gultekin:2009,Schulze:2011,Park:2012}. The inverse ML regression (using Equation~\ref{eq:cond_smu_int}) requires additional information about the distribution function $\Phi_x(x)$ and is therefore more complicated to use. However, in the presence of selection effects on $y$ the inverse ML regression might become preferable, since it can be unaffected by the selection effects, in contrast to the forward ML regression. We will illustrate this below for the $M_\bullet-$bulge relation for active galaxies.

While other applications of this method are possible, in the following we focus on the specific case of studying the $M_\bullet-$bulge relation at high redshift via broad line AGN samples.

\subsection{Application to the $M_\bullet-$bulge relation for broad line AGN samples}
In SW11 we discussed the qualitative influence of sample selection effects on observations of the $M_\bullet-$bulge relation for broad line AGN samples in detail within our framework of the bivariate distribution function.

The $M_\bullet-$bulge relation is given by $\mu = \alpha + \beta s$, with $\mu=\log M_\bullet$ being the logarithmic black hole mass and $s$ is the logarithm of the host galaxy property,  e.g. $\log \sigma_\ast$ or $\log M_\mathrm{bulge}$, and $\alpha$ and $\beta$ are the normalization and the slope of the relation. Thus in the above terminology $x=s$ and $y=\mu$. 
With the parametrization of equation~(\ref{eq:psi}), the bivariate distribution function is 
\begin{equation}
\Psi(s,\mu)=g(\mu \, |\, s)\, \Phi_s(s) \ , \label{eq:psi_agn}
\end{equation}
where $\Phi_s(s,z)$ is the spheroid distribution function and $g(\mu \, |\, s,z)$ is the intrinsic $M_\bullet-$bulge relation. Integrating this bivariate distribution function over $s$ gives the total BHMF.

For an AGN sample the main selection criteria will be the AGN flux. Therefore, the selection function will depend on AGN luminosity and redshift. The AGN luminosity is determined by the black hole mass $M_\bullet$ and the accretion rate, the latter is represented by the Eddington ratio $\lambda=L_\mathrm{bol}/L_\mathrm{Edd}$, i.e. $\log L_\mathrm{bol}=\mu + \log \er +38.1$. 
This implies a selection function $\Omega(\mu,\er,z)$, depending on black hole mass, Eddington ratio and redshift. 

The corresponding distribution function is the bivariate distribution function of black hole mass and Eddington ratio for active black holes $\tilde{\Psi}(\mu,\lambda)$, with its projections the active BHMF and the ERDF. We here make the assumption that the bivariate distribution function of black hole mass and Eddington ratio is separable into an independent  BHMF and  ERDF, i.e. 
\begin{equation}
\tilde{\Psi}(\mu,\lambda,z)=p_\er(\er)\,\Phi_{\bullet,a}(\mu,z) \  ,
\end{equation}
where $p_\lambda(\er,z)$ is the ERDF, normalized to one and $\Phi_{\bullet,a}(\mu,z)$ is the active BHMF.
This assumption is motivated by results on the local bivariate distribution function of $\mu$ and \er\citep{Schulze:2010} and by the observation of an Eddington ratio distribution of AGN that is independent of stellar mass \citep{Aird:2012,Bongiorno:2012}, suggesting also an independence of $\mu$ [but see \citet{Kelly:2013} for evidence of a $\mu$ dependence of the ERDF]. We comment on the consequences of relaxing this assumption below.

The active BHMF is related to the total BHMF via the active fraction, which gives the probability for a black hole to be in an active (type~1 AGN) state as a function of black hole mass and redshift. It is defined by the ratio of active BHMF to total BHMF, $p_\mathrm{ac}(\mu,z)=\Phi_{\bullet,a}(\mu,z)/\Phi_{\bullet,t}(\mu,z)$. The latter is directly linked to the bivariate distribution of $\mu$ and $s$ (Equation~(\ref{eq:psi_agn})).

Following Equation~(\ref{eq:psio}), the multivariate probability distribution $\Psi_\mathrm{o}(s,\mu,\er,z)$ of galaxy property, black hole mass, Eddington ratio and redshift is then given by
\begin{equation}
\Psi_\mathrm{o}(s,\mu,\er,z)= \Omega(\mu,\er,z)\, p_\mathrm{ac}(\mu,z)\, p_\lambda(\er,z)\,g(\mu \, |\, s,z)\, \Phi_s(s,z) \frac{\dd V}{\dd z} \ , \label{eq:multipsiflux_ap}
\end{equation}
where $\Omega(\mu,\er,z)$ is the AGN selection function of the respective sample.
The conditional probability for the forward ML regression then follows as
\begin{eqnarray}
 p(\mu \, |\, s,\er,z) & = & \frac{\Psi_\mathrm{o}(s,\mu,\er,z)}{\int \Psi_\mathrm{o}(s,\mu,\er,z)\, \dd \mu} \\ \nonumber
 & = &  \frac{\Omega\, p_\mathrm{ac}(\mu,z)\,g(\mu \, |\, s,z)}{\int \Omega\, p_\mathrm{ac}(\mu,z)\,g(\mu \, |\, s,z)\,\dd \mu} \ . \nonumber \label{eq:condprob_agn}
\end{eqnarray}
The ERDF $p_\lambda(\er)$ and the galaxy distribution function $\Phi_s(s)$ cancel out and are not required for the fitting. We require only knowledge of the selection function and of the mass dependence of the active fraction to obtain the intrinsic $M_\bullet-$bulge relation from the fit to the conditional probability. The free parameters $\alpha$, $\beta$ and $\sigma$ in the intrinsic relation $g(\mu \, |\, s,z)$ are obtained by minimizing Equation~\ref{eq:Sx}. If the ERDF is mass dependent it will not cancel out and knowledge of the ERDF is additionally required.

Second, the conditional probability for the inverse ML regression is
\begin{eqnarray}
 p(s \, |\, \mu,\er,z) & = & \frac{\Psi_\mathrm{o}(s,\mu,\er,z)}{\int \Psi_\mathrm{o}(s,\mu,\er,z)\, \dd s} \\
 & = & \frac{g(\mu \, |\, s,z)\, \Phi_s(s) }{\int g(\mu \, |\, s,z)\, \Phi_s(s)\, \dd s} \  . \nonumber
\end{eqnarray}
This is identical to the case without AGN selection effects (Equation~\ref{eq:cond_smu_int}), i.e. the conditional probability $p(s \, |\, \mu)$ is not affected by the AGN selection. We already discussed this fact in SW11, but we will use it here to recover the intrinsic relation by the maximum likelihood fit. 
The only quantity that has to be known to recover the intrinsic relation is the galaxy distribution function.

Both fitting options have their pros and cons. While the latter is largely unaffected by AGN selection effects, it is susceptible to the shape of the galaxy distribution function and also more susceptible to the intrinsic scatter in the relation itself. The former does not require any knowledge about $\Phi_s(s)$, but demands a proper understanding of the sample selection and also of the underlying distribution function of AGN or at least the active fraction for the AGN sample. 

Both regression methods constitute an improvement for the determination of the intrinsic $M_\bullet-$bulge relation, compared to a standard regression that does not include knowledge of the selection function or galaxy distribution function, as we will illustrate below.

\subsection{Incorporating measurement uncertainties} \label{sec:mu}
In practice, the situation becomes more complicated, since black hole mass and galaxy property measurements have measurement uncertainties. Virial black hole masses are associated with a non-negligible uncertainty of $\sim0.3-0.4$~dex \citep{Vestergaard:2006,Park:2012}, while also AGN host galaxy properties at high $z$ have significant measurement errors. These lead to second-order dependencies of the conditional probabilities.
In SW11 we already studied the consequences of measurement uncertainties on the bivariate distribution and the $M_\bullet-$bulge relations. Here, we discuss the ramifications of measurement uncertainties on the presented fitting approach.

We assume that the black hole mass estimated by the virial method $\mu_\mathrm{o}$ is drawn from a log-normal probability distribution around the true black hole mass $\mu$, with dispersion $\sigma_\mathrm{vm}$, representing the uncertainty in the virial mass estimate, 
\begin{equation}
 g(\mu_\mathrm{o} \, |\, \mu) = \frac{1}{\sqrt{2 \pi} \sigma_\mathrm{vm}} \exp \left\lbrace - \frac{(\mu_\mathrm{o}-\mu)^2}{2 \sigma_\mathrm{vm}^2} \right\rbrace  \ . \label{eq:gvm}
\end{equation} 
For the galaxy property $s$ we also assume a log-normal error distribution $g(s_\mathrm{o} \, |\, s)$ with dispersion $\sigma_s$.

The bivariate distribution function for bulge property and virial black hole mass is then
\begin{equation}
  \Psi_\mathrm{o}(s_\mathrm{o},\mu_\mathrm{o})= \int g(s_\mathrm{o} \, |\, s)\, g(\mu_\mathrm{o} \, |\, \mu)\, \Psi_\mathrm{o}(s,\mu)\,\dd s \dd \mu \, . \label{eq:psivm}
\end{equation}  
The bivariate distribution function is smoothed out by the measurement uncertainties.
The conditional probability for the forward ML regression is
\begin{multline}
p(\mu_\mathrm{o} \, |\, s_\mathrm{o},\er,z)  =  \\
\frac{ \int \Psi_\mathrm{o}(s,\mu,\er,z)\, g(s_\mathrm{o} \, |\, s)\,g(\mu_\mathrm{o} \, |\, \mu)\, \dd s \dd \mu}{\int \Psi_\mathrm{o}(s,\mu,\er,z)\,  g(s_\mathrm{o} \, |\, s)\, g(\mu_\mathrm{o} \, |\, \mu)\, \dd s \dd \mu \dd \mu_\mathrm{o} }  = \\ 
\frac{\int \Omega\, p_\mathrm{ac}(\mu,z)\,g(\mu \, |\, s,z)\,\Phi_s(s)\,g(s_\mathrm{o} \, |\, s)\,g(\mu_\mathrm{o} \, |\, \mu)\,\dd s \dd \mu}{\int \Omega\, p_\mathrm{ac}(\mu,z)\,g(\mu \, |\, s,z)\,\Phi_s(s)\,g(s_\mathrm{o} \, |\, s)\,\dd s \dd \mu} \ .  \label{eq:condprob_agn_mus}
\end{multline} 
Assuming we know the measurement uncertainties $\sigma_\mathrm{vm}$ and $\sigma_s$, these can be directly incorporated into the fit.
However, we now introduce a dependence on the galaxy distribution $\Phi_s(s)$ into the forward ML regression which was canceled out before. To avoid this dependence, we can make the simplifying assumption $\Phi_s(s)\approx \Phi_s(s_\mathrm{o})$. This approximation is justified for a small measurement error in $s$ and in particular in the flat part of the galaxy distribution function. It will  break down for high stellar velocity dispersion or high stellar mass, i.e. at the exponential decline of the galaxy distribution function. It can be applied to samples dominated by lower mass galaxies.
Equation~\ref{eq:condprob_agn_mus} then simplifies to
\begin{equation}
p(\mu_\mathrm{o} \, |\, s_\mathrm{o},\er,z)  \approx 
\frac{\int \Omega\, p_\mathrm{ac}(\mu,z)\,g(\mu \, |\, s,z)\,g(s_\mathrm{o} \, |\, s)\,g(\mu_\mathrm{o} \, |\, \mu)\,\dd s \dd \mu}{\int \Omega\, p_\mathrm{ac}(\mu,z)\,g(\mu \, |\, s,z)\,g(s_\mathrm{o} \, |\, s)\,\dd s \dd \mu} \ .  \label{eq:condprob_agn_mus_simp}
\end{equation}
We will refer to both options as forward ML regression with measurement errors.

The conditional probability for the inverse ML regression is given by
\begin{multline}
p(s_\mathrm{o} \, |\, \mu_\mathrm{o},\er,z)  =  \\
\frac{ \int \Psi_\mathrm{o}(s,\mu,\er,z)\, g(s_\mathrm{o} \, |\, s)\,g(\mu_\mathrm{o} \, |\, \mu)\, \dd s \dd \mu}{\int \Psi_\mathrm{o}(s,\mu,\er,z)\,  g(s_\mathrm{o} \, |\, s)\, g(\mu_\mathrm{o} \, |\, \mu)\, \dd s \dd \mu \dd s_\mathrm{o} }  = \\ 
\frac{\int \Omega\, p_\mathrm{ac}(\mu,z)\,g(\mu \, |\, s,z)\,\Phi_s(s)\,g(s_\mathrm{o} \, |\, s)\,g(\mu_\mathrm{o} \, |\, \mu)\,\dd s \dd \mu}{\int \Omega\, p_\mathrm{ac}(\mu,z)\,g(\mu \, |\, s,z)\,\Phi_s(s)\,g(\mu_\mathrm{o} \, |\, \mu)\,\dd s \dd \mu} \ .  \label{eq:condprob_agn_smu}
\end{multline} 
For this case, the dependence of the conditional probability on the active fraction and on the selection function is not canceled out as it was the case without measurement uncertainty in $\mu$. This effectively prevents ignoring the AGN selection function for the fit of the $M_\bullet$-bulge relations for observational samples.

We can again derive an approximation to the probability distribution by making the simplifying assumptions $p_\mathrm{ac}(\mu)\approx p_\mathrm{ac}(\mu_\mathrm{o})$ and $\Omega(\mu)\approx \Omega(\mu_\mathrm{o})$, for which Equation~(\ref{eq:condprob_agn_smu}) reduces to
\begin{equation}
p(s_\mathrm{o} \, |\, \mu_\mathrm{o},\er,z)  \approx  
\frac{\int g(\mu \, |\, s,z)\,\Phi_s(s)\,g(s_\mathrm{o} \, |\, s)\,g(\mu_\mathrm{o} \, |\, \mu)\,\dd s \dd \mu}{\int g(\mu \, |\, s,z)\,\Phi_s(s)\,g(\mu_\mathrm{o} \, |\, \mu)\,\dd s \dd \mu} \ .  \label{eq:condprob_agn_smu_simp}
\end{equation}
For this approximation no knowledge of the AGN selection function and the active fraction is required anymore.
The assumption on the selection function depends on the details of the sample selection, but in general will be true in most cases, unless the object is close to the selection threshold (e.g. to the flux limit of the survey). The assumption on the active fraction requires an active fraction with a weak or no black hole mass dependence. This is, for example, the case at $z\sim1.5$, as discussed in Section~\ref{sec:app_z15}, but not in the local universe \citep{Schulze:2010}. Therefore, in practice some idea about the active fraction is required in any case to judge the validity of this approximation.  

We will refer to these maximum likelihood fits as inverse ML regression with measurement errors.

We here presented several options to fit for the underlying relation using a given sample with measurement uncertainties. Which of these is the best choice for the specific sample will depend on the details of the sample selection and our knowledge of this selection function and the galaxy distribution function.
Formally, the most precise results are obtained using the conditional probabilities for either the forward or inverse ML regression given by Equations~\ref{eq:condprob_agn_mus} and  \ref{eq:condprob_agn_smu}. This requires knowledge of the selection function, the active fraction and of the galaxy distribution function. To restrict the assumptions for the respective fit, we can also use the approximations given by  Equations~\ref{eq:condprob_agn_mus_simp} and  \ref{eq:condprob_agn_smu_simp}. While not precise, this approach can be justified if the selection function and/or the underlying distributions are only poorly known.

\begin{figure*}
\centering 
\includegraphics[width=18cm,clip]{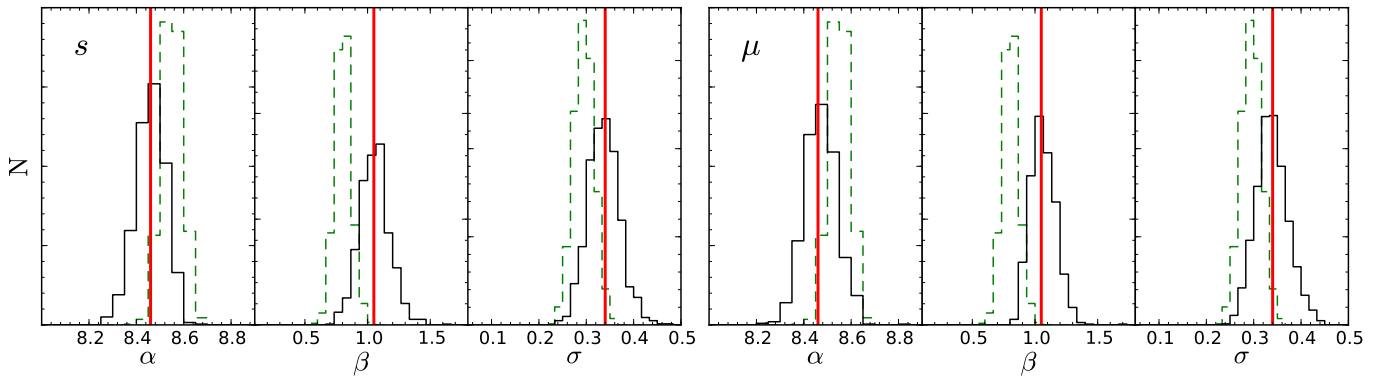} 
\caption{
Demonstration of biases due to selection effects and their removal by our ML fitting approach. For a set of Monte Carlo simulations we assumed the 'true' $\mbh$-bulge relation at $z=1.5$ to be described by a normalization $\alpha$, a slope $\beta$, and an intrinsic scatter $\sigma$; these input values are marked by the vertical red lines in each panel. For 1000 simulated random samples of 100 objects each,  with a flux limit of $i < 22$~mag, the panels show the best-fitting parameters from simple linear regression (green dashed histograms) and from our ML regression method (black histograms -- left subpanels: forward regression, right subpanels: inverse regression). The horizontal offsets between the red lines and the centroids of the dashed histograms indicate clearly the presence of biases, while the estimates using our ML regression method are largely unbiased.}
\label{fig:mcfit}
\end{figure*}

\begin{figure*}
\centering 
\includegraphics[width=18cm,clip]{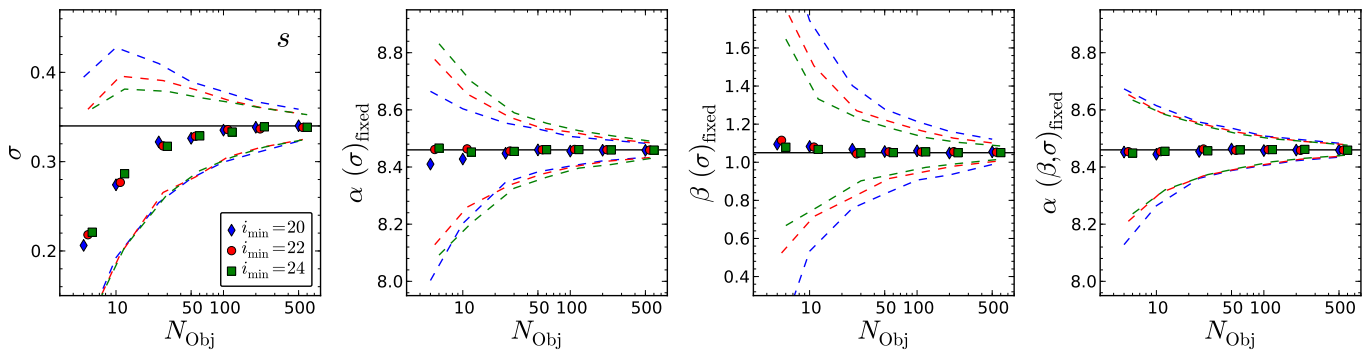} 
\caption{
Recovery of intrinsic relation parameters by using our forward ML regression method, accounting for selection effects. The Monte Carlo simulations are the same as for Fig.~\ref{fig:mcfit}, but for a range of sample sizes and flux limits. Each panel shows median (symbols) and 68\% confidence intervals (dashed lines) for the best-fitting results of normalization $\alpha$, slope $\beta$, and intrinsic scatter $\sigma$, respectively. The horizontal black line indicates the input value. Both $\alpha$ and $\beta$ can be recovered without any systematic bias even from small samples, whereas $\sigma$ tends to be somewhat underestimated unless the sample is large enough.}
\label{fig:statsX}
\end{figure*}

\begin{figure*}
\centering 
\includegraphics[width=18cm,clip]{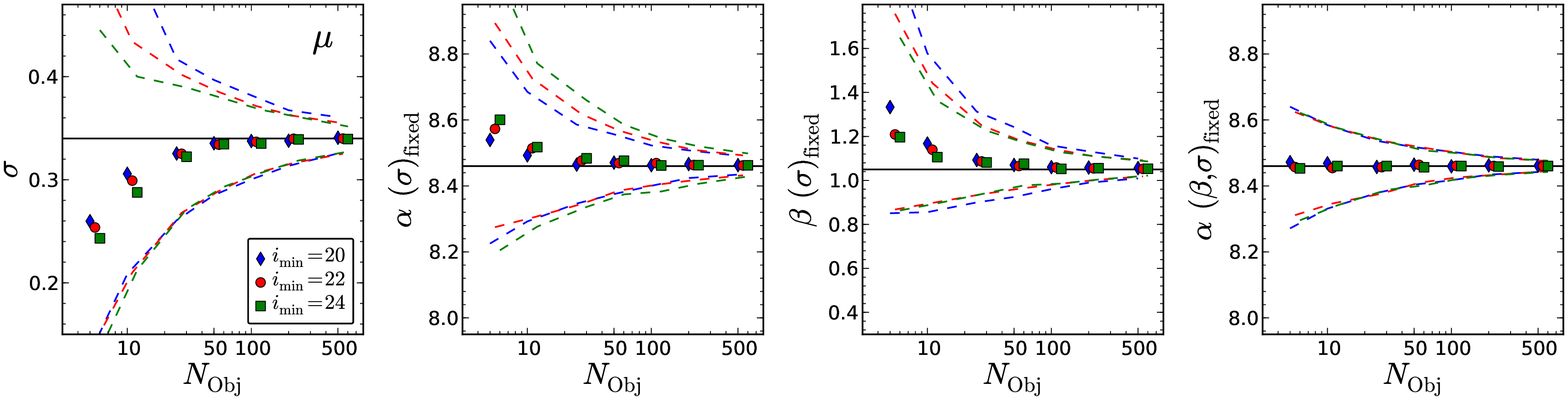} 
\caption{Same as Fig.~\ref{fig:statsX}, but for an inverse ML regression. Again this verifies the reliability of the method to reconstruct the intrinsic relation.
} 
\label{fig:statsY}
\end{figure*}

\section{Monte Carlo simulations} \label{sec:predict}
To illustrate the method and test its reliability  we carried out Monte Carlo simulations of the $M_\bullet-M_\ast$ relation. For our simulations we chose as representative redshift $z=1.5$. For the galaxy spheroid mass function we assume the estimate given in SW11, which is based on the stellar mass function from \citet{Fontana:2006}, applying a rough correction for the bulge-to-total ratio to translate it into a spheroid mass function. We set up the simulations by first drawing a large number of objects from the spheroid mass function at $z=1.5$. Black hole masses are attributed according to the $M_\bullet-M_\mathrm{bulge}$ relation from \citet{McConnell:2013} with log-normal intrinsic scatter $\sigma=0.34$. We here assumed a constant active fraction,  consistent with recent results on the active BHMF at $z\sim1.5$ \citep[][Schulze et al. in prep.]{Nobuta:2012,Kelly:2013} and empirical models \citep{Merloni:2008,Shankar:2009}. 
For the active black holes an Eddington ratio was drawn from the $z\sim1.5$ ERDF from \citet{Nobuta:2012}. Black hole mass and Eddington ratio define the bolometric AGN luminosity. 
We convert $L_\mathrm{bol}$ into $i$-band magnitude using the AGN spectral energy distribution (SED) from \citet{Richards:2006a} and the $K$-correction from \citet{Richards:2006b}. We applied three flux limits to the simulated samples, $i_\mathrm{min}$=20, 22 and 24~mag, respectively, and generated samples of various sizes. For each flux limit and sample size we generated 1000 Monte Carlo realizations. 

For simplicity, we first discuss the results without measurement errors. We will discuss the effect of adding measurement errors further below. Each simulation realization was fitted by a forward and an inverse ML regression. Furthermore, for each regression we fitted the realizations with a different set of free parameters. We first kept all three parameters ($\alpha$, $\beta$, $\sigma$) free. Secondly we fixed $\sigma$ and finally we also fixed $\beta$ to the input value, leaving only the normalization as free parameter. Additionally, we fitted each realization with a simple linear regression, i.e. not including the selection effects, both with $\mu$ and with $s$ as dependent variable.

\begin{figure*}
\centering 
\includegraphics[width=16cm,clip]{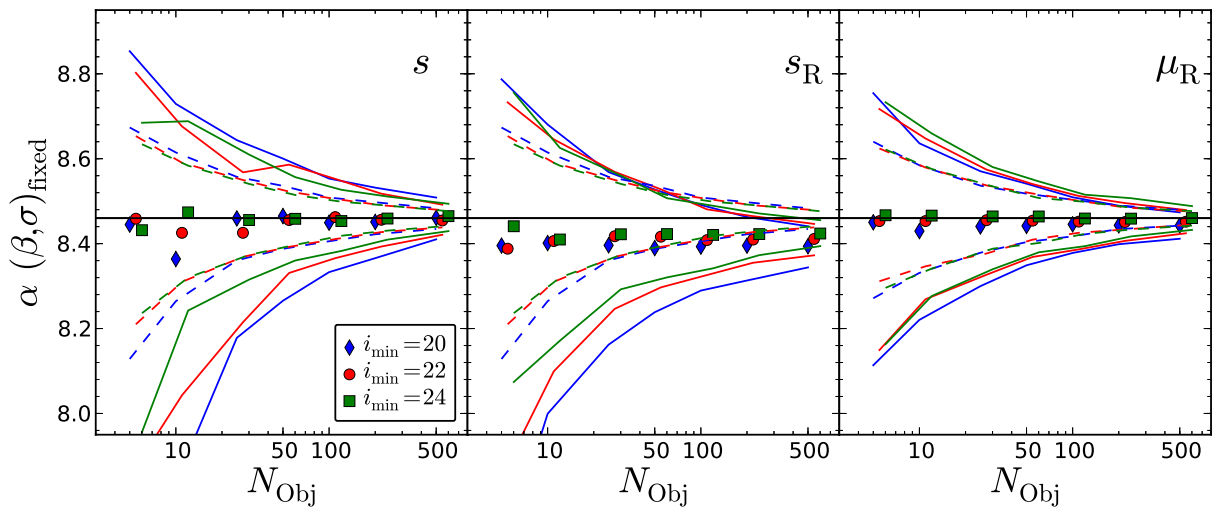}
\caption{
Reliability of our fitting method when measurement errors on the black hole mass and spheroid mass are present. We again show the best-fitting normalization $\alpha$ (for fixed slope and intrinsic scatter) from Monte Carlo simulations at $z=1.5$ for different flux limits, but now with measurement errors added. The different panels show different variants for the fitting, as explained in the text.
Left panel:  using a forward ML regression with measurement errors. Middle panel: using a forward ML regression, but without accounting for measurement errors in the fitting of the relation. Right panel: using an inverse ML regression without accounting for measurement errors. The filled symbols show the median value for different applied magnitude limits, the solid lines show the 1$\sigma$ confidence region for the simulations with measurement errors, while the dashed line shows the 1$\sigma$ confidence region of the fit when no measurement errors were added to the simulated data.
}
\label{fig:stats_error}
\end{figure*}

In Fig.~\ref{fig:mcfit}, we illustrate the strength of our proposed fitting approach (shown as black histograms, left for the forward ML regression and right for the inverse ML regression), compared to a simple linear regression (shown as green histograms). We use 1000 realizations of a sample of $N=100$ objects at $z=1.5$ with $i<22$~mag. While the dispersion on the fitting parameters is increased, the bias present in the simple regression approach is removed. In this example we leave all three parameters free in the fit. 

In Fig.~\ref{fig:statsX} we show the median and 68\% confidence limits of the 1000 realizations for a forward ML regression for different flux limits and sample sizes. The magnitude limits of $i=20,  22, 24$~mag correspond to bolometric luminosity limits of $\log L_\mathrm{bol}=46,  45, 44.3$, respectively.
The left panel shows that  we systematically underestimate $\sigma$ for a small sample size when keeping it as a free parameter. This is due to the fact that the fitting procedure determines the \emph{sample} standard deviation, which is known to be an underestimate of the \emph{population} standard deviation. This underestimate for small sample sizes is already present in the conventional fit without including the selection function. Since the intrinsic scatter is an important parameter, when the selection effects are included this underestimate is enhanced. Also the best-fitting solutions for the normalization $\alpha$ and the slope $\beta$ will be biased. Therefore, we suggest to fix the intrinsic scatter for small sample sizes, $N_\mathrm{Obj}\lesssim50$, to avoid this bias. In the following we fix the intrinsic scatter  $\sigma$ in the maximum likelihood fit. In this case, the slope $\beta$ and normalization $\alpha$ are recovered unbiased irrespective of sample size, as shown in the other three panels of Fig.~\ref{fig:statsX}. Increasing the sample size mainly reduces the statistical errors, shown by the dashed lines. Fixing also the slope and keeping the normalization as only free parameter further reduces the statistical error, while the input is recovered in the mean. We further see that the error is basically independent of the flux limit. However, this assumes perfect knowledge of the selection function and underlying distribution. Deeper samples are less susceptible to uncertainties in these and therefore come closer to the statistical errors shown here.

In Fig.~\ref{fig:statsY} we show the same for the inverse ML regression. Keeping the intrinsic scatter as a free parameter also leads to an underestimate for small sample sizes and we again suggest to fix it for small samples, $N_\mathrm{Obj}\lesssim50$, to avoid a bias in the slope and normalization. But even when fixing $\sigma$ we find in our simulations a small bias in $\alpha$ and $\beta$ for very small samples,  $N_\mathrm{Obj}\lesssim10$. However, for such small samples the errors on the best fit are also large. A joint determination of slope and normalization of the relation requires a reasonably large sample. When fixing the slope $\beta$ and only determining the normalization $\alpha$, an unbiased result with small error can be achieved even for relatively small sample sizes, as demonstrated in the right panel of Fig.~\ref{fig:statsY}.

We now investigate the consequences of adding measurement errors in the virial black hole masses and in the host galaxy properties to the simulations. We here test our prescription for including these uncertainties, as presented in section~\ref{sec:mu} and compare them to a regression without measurement errors. 
We added a log-normal random error of dispersion $\sigma_\mathrm{vm}=0.3$~dex to the black hole mass and a log-normal error of dispersion $\sigma_s=0.3$~dex to the spheroid mass of our Monte Carlo realizations at $z=1.5$. For each realization we fixed $\beta$ and $\sigma$ and only fit for the normalization $\alpha$.  We fitted these with the forward ML regression with measurement errors, i.e. using Equation~\ref{eq:condprob_agn_mus}. The difference to the inverse ML regression with measurement errors is marginal, in particular when the errors in $\mu$ and $s$ are of the same magnitude. To test the importance of the full correction for measurement errors, we also fitted the same realisations with the forward and inverse ML regression without measurement errors.
 
The results are shown in Fig.~\ref{fig:stats_error}. We show the median (filled symbols), the 68\% confidence limits (solid lines) and also the 68\% confidence limits for the realizations with zero error in $\mu$ and $s$ (dashed lines).  In the left panel the result for the forward ML regression with measurement errors is presented.
The input value is recovered in the median, while the 1$\sigma$ confidence region (solid lines) is increased by an approximately constant factor of $\sim2$ for our simulation, compared to the case without measurement errors (dashed lines). The fitting method provides unbiased results if selection function, underlying distributions and the value of the measurement error are all known.
In the middle and right panels of Fig.~\ref{fig:stats_error} we show the results of applying the forward ML regression and inverse ML regression without measurement errors to the simulations that have errors added. While in the forward ML regression the zero-point $\alpha$ is slightly biased towards a lower value, the inverse ML regression results are fully consistent with the input value, only showing an increase in the 1$\sigma$ confidence range. Thus, not fully including the measurement errors has only a small, second order effect on the best-fitting solution.

The Monte Carlo simulations verify the reliability of the presented maximum likelihood fitting approach to recover the intrinsic $M_\bullet-M_\ast$ relation in the median from a sample affected by sample selection effects. A simple linear regression would provide a biased result in these cases. The method is still reliable if there are measurement uncertainties in the black hole mass and galaxy property. While the most precise results are obtained when the uncertainty is well known and taken into account, over- or underestimating or even ignoring the measurement uncertainty will only have a small effect on the recovery of the intrinsic relation.

\section{Application to observational studies} \label{sec:application}
We now  apply our method to a few observational studies from the literature that investigated the evolution in the $M_\bullet-$bulge relations. This is meant for illustration of the methodology and constraining power of the approach. An exhaustive investigation of the literature on evolution in the $M_\bullet-$bulge relations in consideration of selection effects is beyond the scope of this paper. We note that such an investigation would also be hampered by the fact that many previous literature studies did not use well-defined samples, or at least samples for which a selection function (however complicated) could be defined.

\begin{figure}
\centering 
\resizebox{\hsize}{!}{\includegraphics[width=17cm]{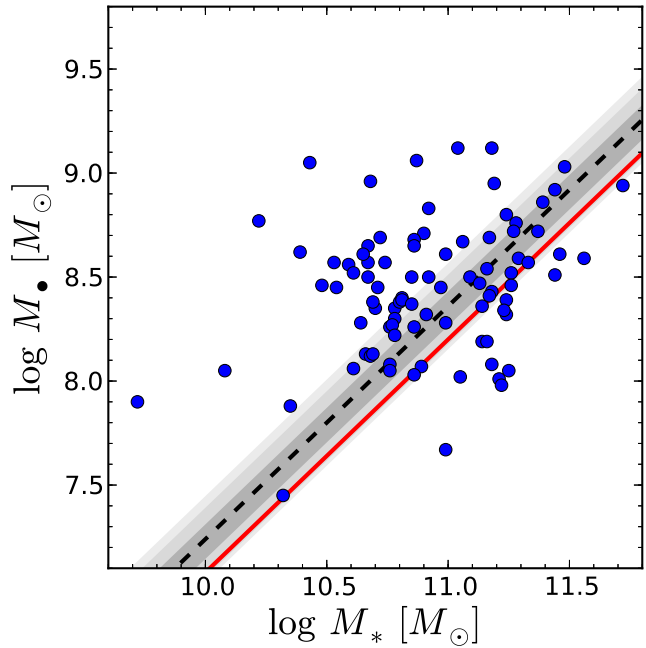}}
\caption{Redshift evolution fit to the AGN sample from \citet{Merloni:2010}. We fitted their data (blue circles) to the conditional probability distributions  $p(y_\mathrm{o} \, |\, x_\mathrm{o},\er,z)$, determining the evolution term $\gamma$. The red solid line shows the local relation from \citet{McConnell:2013}, and the black dashed line gives the best-fitting relation at the mean redshift of the sample. The grey regions are the $1\sigma$, $2\sigma$ and $3\sigma$ confidence regions.}
\label{fig:fitz_merloni10}
\end{figure}

\subsection{The $M_\bullet-M_\ast$ relation at $z\sim1.5$} \label{sec:app_z15}
First, we reinvestigated the QSO sample from \citet{Merloni:2010}. They studied the $M_\bullet-M_\ast$ relation for 89 broad line AGN from zCOSMOS in the redshift range $1.06<z<2.19$ with estimated stellar masses for the host galaxies (including upper limits). The selection function is to first order defined by an $I_\mathrm{AB}$-band flux limit of 22.5~mag.

To fit their $M_\bullet-M_\ast$ relation we need additional information about the underlying distribution functions. We use the same assumptions as above for the Monte Carlo realizations, i.e. the stellar mass function from \citet{Fontana:2006}, and a constant active fraction.
To investigate the degree of evolution, we assume a redshift evolution in the zero-point $a(z)= a_0+\gamma \log (1+z)$. For the local relation we use first the relation by \citet{Haering:2004}, with an intrinsic scatter of $\sigma=0.3$, consistent with the work by \citet{Merloni:2010}, and second the recent results by \citet{McConnell:2013}, with $\sigma=0.34$.  Upper limits are incorporated as described in Appendix~\ref{sec:upperlimits}.

For the forward ML regression (using Equation~\ref{eq:condprob_agn_mus}), assuming measurement errors of $0.2$~dex in the stellar mass and $0.3$~dex in the black hole mass estimates, we get $\gamma=0.38^{+0.20}_{-0.21}$, using the \citet{Haering:2004} local relation. Using the relation from \citet{McConnell:2013} as zero-point, we find $\gamma=-0.35^{+0.22}_{-0.23}$ (i.e. formally a negative evolution).  For the inverse ML regression and when using approximations for the probability distribution (Equations~\ref{eq:condprob_agn_mus_simp} and  \ref{eq:condprob_agn_smu_simp}) we get consistent results. The null hypothesis of no evolution in the \mbhbu relation for both cases lies within $2-3\sigma$ of our results, thus it is not rejected with statistical significance. The best fit relation, using the \citet{Haering:2004} local relation, is shown in Fig.~\ref{fig:fitz_merloni10}.

Note that the given errors are statistical errors from the $\Delta S$ contour, while uncertainties in the assumptions for the zero-point, the flux limit  and the distribution functions are not included. This implies that the uncertainty estimates above are rather optimistic, since they only include statistical uncertainties and not additional uncertainties in the underlying assumptions. The likelihood for the null hypothesis of no evolution is then even higher than the $2-3\sigma$ mentioned above when those uncertainties are also included in the confidence determination. 

To get an estimate for the effect of uncertainties in the underlying distribution function, we varied the assumed stellar mass function from \citet{Fontana:2006} within their quoted uncertainties. Furthermore, we varied the bulge-to-total ratio between our default correction and no correction at all, to investigate the importance of the transformation of the galaxy mass function to a spheroid mass function. The uncertainty in the stellar mass function adds an uncertainty of $\pm0.05$ to the best-fitting $\gamma$, while not applying an bulge-to-total correction increases the best-fitting solution by $0.04$.

Even more important is  a precise calibration of the  local zero-point, as it is illustrated by the opposite trends of evolution found for the local relations by \citet{Haering:2004} and \citet{McConnell:2013}, which is mainly caused by the increase in the zero-point of the $M_\bullet-M_\mathrm{Bulge}$ relation of \citet{McConnell:2013}. 

While the local zero-point of the $M_\bullet-M_\mathrm{Bulge}$ relation increased in the most recent revised determinations of the relation \citep{McConnell:2013,Kormendy:2013}, the $f$ factor and thus the calibration to the AGN samples is not strongly affected by this revision \citep{Park:2012,Woo:2013}. Using this revised $M_\bullet-M_\mathrm{Bulge}$ relation significantly reduces the apparent trend of positive evolution, a fact recently noted by \citet{Kormendy:2013}. Combined with the correction for the bias caused by sample selection this even suggests or is at least consistent with a negative evolution of the $M_\bullet-M_\mathrm{Bulge}$ relation where galaxy bulges grow first and the black holes have to catch up.

In summary, we see no statistically significant evidence for evolution in the \mbhbu relation based on this data set, consistent with our more qualitative analysis in SW11.

\subsection{High redshift QSOs}
Studies of the most luminous QSOs, observed at high redshift found the largest apparent offsets from the $M_\bullet-$bulge relations, by a factor of 10 or more \citep{Walter:2004,Maiolino:2007,Ho:2007,Riechers:2008,Schramm:2008,Wang:2010,Targett:2012}. At first glance this seems to provide the most convincing evidence for strong evolution in the $M_\bullet-$bulge relations. However, these luminous QSO studies are only sampling from the  bright end of the luminosity function. For this kind of objects the bias caused by selection effects is most severe, as discussed in SW11. Furthermore, current sample sizes at $z>3$ are small, inhibiting firm statistical conclusions.

We now endeavor to test quantitatively if the strong apparent offset found in these studies provides any statistically significant evidence for an intrinsic offset, once we account for sample selection. We focus on the highest redshift regime currently studied $z\sim6$ and the constraints that can be drawn on evolution in the $M_\bullet-$bulge relations. 

We used the sample from \citet{Wang:2010}, and restricted it to the six objects selected from the Sloan Digital Sky Survey (SDSS) $z>5.7$ quasar survey \citep{Fan:2000,Fan:2001,Fan:2003,Fan:2006}, which approximately might be considered  a well-defined sample. For this sample we assume a selection function given by the survey magnitude limit of $z^*<20.2$~mag. Black hole masses are for some of the objects estimated from the broad lines and for the rest from the bolometric luminosity, assuming Eddington limited accretion. Bulge masses are derived from dynamical mass measurements (we assume an average inclination angle of $40^\circ$), and stellar velocity dispersions were estimated from the width of molecular CO line emission. There are still significant systematic uncertainties associated with these measurements and how they relate to galaxy masses and velocity dispersion measured for local galaxies. However, we here only discuss the  statistical aspects of the sample.

The dynamical range in black hole mass and AGN luminosity covered by this sample is small and there is basically no apparent correlation between $M_\bullet$ and either $\sigma_\ast$ or $M_\mathrm{dyn}$. 
We therefore fixed the slope and intrinsic scatter and only fit for an evolution in the normalization, again parametrized by a parameter $\gamma$.

For the stellar mass function we assume the mass function for star forming galaxies at $z\sim5$ from \citet{Lee:2012}. We again assume a constant active fraction.  
Using the forward regression with measurement errors we find $\gamma^{M_\ast}<1.10$ at $1\sigma$ confidence, with a best fit of $\gamma^{M_\ast}=-0.02$ for the \mbhbu relation, using the local relation from \citet{Haering:2004}.  We also varied the mass function according to their uncertainties, as given in \citet{Lee:2012}. The best-fitting $\gamma^{M_\ast}$ ranges from $-0.74$ to $0.59$ with an $1\sigma$ confidence interval $[-3.24, +1.63]$. For the local relation from \citet{McConnell:2013} as zero-point we find a best fit of $\gamma^{M_\ast}=-0.58$ with $\gamma^{M_\ast}<0.71$ at $1\sigma$ confidence. At any rate the inferred value of $\gamma$ is always consistent with zero.

\begin{figure}
\centering 
\resizebox{\hsize}{!}{\includegraphics[clip]{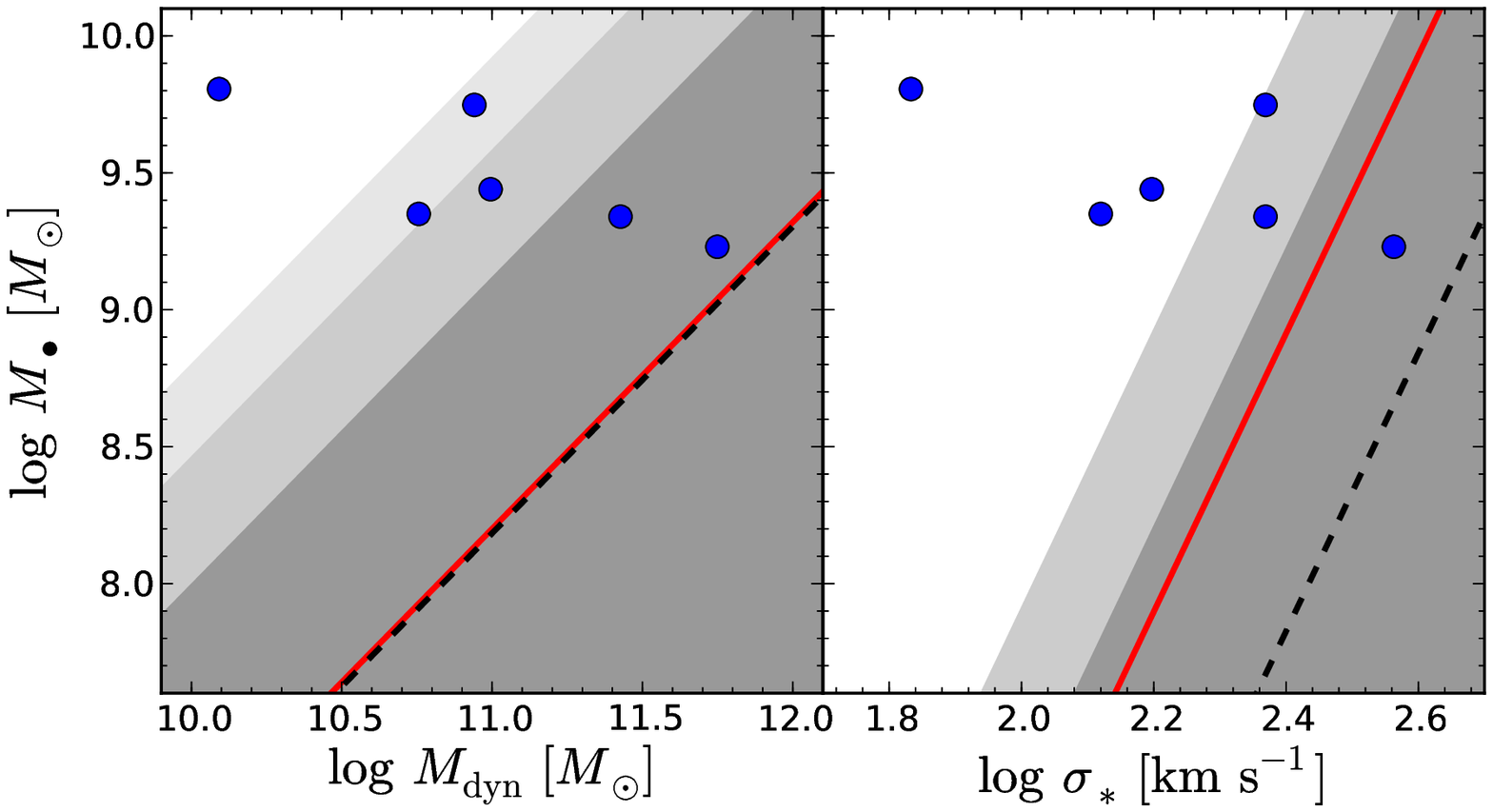}}
\caption{Redshift evolution fit of the \mbhbu relation (left panel) and \mbhsig relation (right panel) to the main SDSS QSO sample given in \citet{Wang:2010}. We used a forward regression, accounting for selection effects and including measurement errors. The blue circles show the data, the red solid line is the local relation, the dashed line is the best-fitting relation at the mean redshift of the sample. The grey areas indicate confidence regions of $1\sigma$, $2\sigma$ and $3\sigma$, respectively.}
\label{fig:fitz_wang10}
\end{figure}

Assuming a SVDF is more problematic. The SVDF is well determined for $z=0$ \citep{Sheth:2003,Chae:2010} and there are estimates out to $z\approx1.5$ \citep{Chae:2010,Bezanson:2011}, but it is observationally unconstrained at higher redshifts. On the other hand, it is not clear at all to what degree the CO line width traces the stellar velocity distribution at these redshift. We here simply try to bracket our ignorance about the SVDF at $z\sim6$ and investigate their consequences on the results for the $M_\bullet-\sigma_\ast$ relation. Note that for the correction only the shape of the distribution function is important, in particular the location of the Schechter function cutoff, not its normalization.

We started from the evolution of the stellar mass function between $z=0$ and $z\sim5$, comparing the mass functions from \citet{Bell:2003} and \citet{Lee:2012}. They show a decrease of the break of the mass function of $\sim0.6$~dex. Assuming virialized systems, we can convert this into a shift for the SVDF. However, this also requires a parameterization of the size evolution of galaxies.  At $z\sim2$ massive galaxies are more compact by a factor of $3-5$, compared to local galaxies of similar mass \citep[e.g.][]{Daddi:2005,Trujillo:2007,vanDokkum:2008}. Assuming similar compact galaxies at $z\sim5-6$ this approximately balances out the evolution in the mass function, i.e. no evolution in the break of the SVDF would be required, consistent with observations out to $z=1.5$ \citep{Bezanson:2011}. We use the local SVDF from \citet{Chae:2010} and converted it to a rough guess for the shape of the SVDF at $z\approx6$, testing two extreme evolution scenarios: (1) no effective evolution in the SVDF, due to an approximate balance between mass evolution and size evolution of galaxies, (2) no size evolution, leading to a decrease in the break of the SVDF, consistent with the mass function evolution.

For the first case we find $\gamma^{\sigma_\ast}<0.17$ at $1\sigma$ confidence, with a best fit of $\gamma^{\sigma_\ast}=-1.44$, compared to the local $M_\bullet-\sigma_\ast$ relation of \citet{McConnell:2013}, while the second case gives $\gamma^{\sigma_\ast}<1.76$ at $1\sigma$ confidence, with a best fit of $\gamma^{\sigma_\ast}=0.25$. In Fig.~\ref{fig:fitz_wang10}, we show the data and the best-fitting results for the $M_\bullet-M_\mathrm{dyn}$ and the $M_\bullet-\sigma_\ast$ relation.

While the uncertainties in the parameters are large, due to both the small number statistics and the uncertainty in the underlying distributions, the results are fully consistent with no evolution in the $M_\bullet-$bulge relations.
Thus, even while these $z\sim6$ QSOs show large apparent offsets from the local relation, this cannot be taken as evidence for intrinsic evolution in the $M_\bullet-$bulge relations up to high $z$. The apparent offset is consistent with being fully caused by  AGN sample selection effects. Even negative evolution in the $M_\bullet-$bulge relations at $z\sim6$ is fully consistent with these data. 

Besides current methodical challenges in determining black hole masses and galaxy properties at $z\sim6$, the significance on the $M_\bullet-$bulge relation at high $z$ is currently limited by small sample sizes and their bright flux limits. Ongoing and future surveys, like UKIDSS, Pan-STARRS, VISTA, SuMIRe/Hyper-Suprime Cam, LSST will significantly increase the number of known $z\sim6$ QSOs  and future sub-mm observations of their host galaxies, in particular using the Atacama Large Millimeter/submillimetre Array (ALMA), will open a new window to the study of their host galaxies and are able to obtain dynamical masses from molecular line emission for a reasonably large sample of high-$z$ QSOs \citep{Wang:2013,Willott:2013}.

\subsection{Current constraints on evolution of the $M_\bullet-$bulge relation}
While we here do not aim at a full investigation of the literature on the evolution of the $M_\bullet-$bulge relation, we want to summarize  the current constraints we have on evolution or non-evolution. We base this on samples that are representative for the best data at their respective redshift, in terms of depth and well defined sample selection. We include the samples from \citet{Merloni:2010} and \citet{Wang:2010} discussed above. Additionally, we augmented this by the samples from \citet{Cisternas:2011} at $z\sim0.5$ and \citet{Targett:2012} at $z\sim4$.

\citet{Cisternas:2011} studied a sample of 32 AGN at $0.3 < z < 0.9$, selected from \textit{XMM}-COSMOS and report basically no offset from the local relation from \citet{Haering:2004} for their sample. Using the \textit{XMM}-COSMOS soft X-ray flux limit, the stellar mass function from \citet{Fontana:2006}, and assuming a constant active fraction, we find an insignificant bias for this sample. Thus, our inferred intrinsic relation is also fully consistent with the local relation from \citet{Haering:2004}. However, compared to the local relation from \citet{McConnell:2013} the sample is offset by $-0.25$~dex. This might indicate a systematic offset in the normalization of the virial black hole mass estimates, for example due to selection effects (SW11). In the following we use the $M_\bullet-$bulge relation from \citet{Haering:2004} as local zero-point as a conservative constraint on possible positive evolution.

\citet{Targett:2012} could resolve the host galaxies for two luminous QSOs at $z=4.16$, selected from the SDSS, with an additional selection criteria $M_i<-28$~mag. To account for the selection effects, we applied this absolute magnitude limit and  assume the mass function for star-forming galaxies at $z\sim4$ from \citet{Lee:2012} and a constant active fraction. The best-fitting result for the forward regression is $\gamma=0.21^{+1.36}_{-2.50}$.

\begin{figure}
\centering 
\resizebox{\hsize}{!}{\includegraphics[clip]{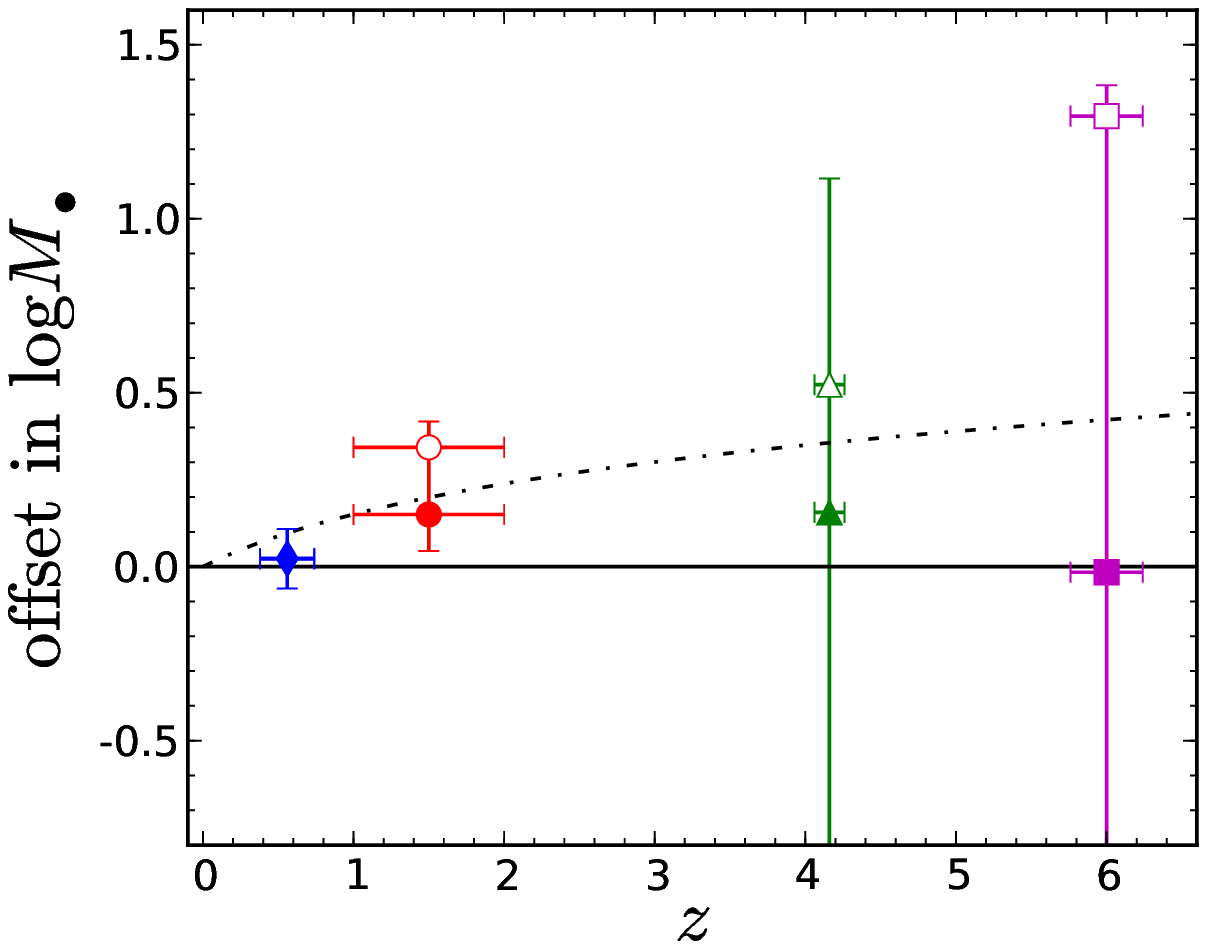}}
\caption{Redshift evolution of the inferred \textit{intrinsic} $M_\bullet-$bulge relation (filled symbols), compared to the apparent relation (open symbols). We here studied the samples by \citet{Cisternas:2011}  at $<z>=0.56$ (blue diamond), \citet{Merloni:2010} at $<z>=1.5$ (red circle), \citet{Targett:2012} at $<z>=4.2$ (green triangle) and \citet{Wang:2010} at $z\sim6$ (magenta square). The error bars indicate the statistical uncertainty of the best fit, including uncertainties in the underlying distributions. The dashed dotted line indicates  a evolution model with $\gamma=0.5$.}
\label{fig:bhevo_intrinsic}
\end{figure}

In Fig.~\ref{fig:bhevo_intrinsic} we summarize our study of the \textit{intrinsic} evolution in the $M_\bullet-$bulge relation for these four samples. The \textit{apparent} offset from the local relation is shown by the open symbols, while our inferred \textit{intrinsic} offset is given by the filled symbols. While there is an apparent evolution trend, accounting for selection effects we find  an \textit{intrinsic} offset that is fully consistent with a non-evolving  $M_\bullet-$bulge relation out to $z\sim6$. However, we note that at $z>2$ the uncertainties on the $M_\bullet-$bulge relation are still substantial and future work in particular in this redshift regime is necessary to obtain robust results. For reference we also indicate  the case of $\gamma=0.5$, predicted by some numerical simulations \citep{DiMatteo:2008,Booth:2010}, by the dashed dotted line in Fig.~\ref{fig:bhevo_intrinsic}. Current data are not able to robustly distinguish between a $\gamma=0.5$ scenario and a non-evolving $M_\bullet-$bulge relation.

\section{Conclusions} \label{sec:conclusions}
While the redshift evolution of the $M_\bullet-$bulge relation provides important constraints on black hole-galaxy co-evolution, its observational study is hampered by severe selection effects.

We presented a novel fitting method to the $M_\bullet-$bulge relations and their evolution. This method corrects for the bias on the $M_\bullet-$bulge relation caused by sample selection and is able to recover the intrinsic relation in the fitting. It does so by incorporating the selection function of the observational sample and the underlying distribution function into the conditional probability distribution of black hole mass at given bulge property or vice versa. This probability distribution is derived from the bivariate distribution function of  $M_\bullet$ and bulge property. The intrinsic relation is reconstructed via a maximum likelihood fit of this  conditional probability distribution to the data. 

Since we are not modeling or fitting the full bivariate distribution function but only the conditional probability distribution, we can reduce the set of prior knowledge or assumptions  to correct for the selection bias, compared to our discussion in SW11.
We presented two routes to fit the conditional probability: (1) using the probability of finding $M_\bullet$ at a given bulge property (forward regression), and (2) using the probability of finding a specific value for the bulge property at a given $M_\bullet$ (inverse regression). For the first case it is essential to properly understand the selection of the sample and have a fair estimate of the black hole mass dependence of the active fraction. For the second case the bulge distribution function and the intrinsic scatter of the $M_\bullet-$bulge relation need to be known. When measurement errors on  $M_\bullet$ and the bulge property cannot be ignored, all of these prior assumptions have to be known for a full correction.

We extensively tested the fitting method on Monte Carlo simulations and verified the reliability of the approach to recover evolution in the normalization unbiased.
We also illustrated the method by applying it to literature data from $z\sim0.5$ to $z\sim6$. The sample of \citet{Merloni:2010} at $z\sim1.5$ already puts significant constraints on the redshift evolution of the $M_\bullet-M_\ast$ relation. Assuming the local relation from \citet{Haering:2004}, we can constrain the evolution  in the $M_\bullet-M_\ast$ relation  to $\gamma<0.9$ out to redshift $z\sim2$ at $3\sigma$ confidence. This is in general agreement with our conclusions in SW11 and with recent results at lower redshifts \citep[e.g.][]{Cisternas:2011, Zhang:2012,Salviander:2013, Schramm:2013}. Taking these empirical clues together, it becomes apparent that there is currently no evidence for significant positive evolution in the $M_\bullet-$bulge relation out to $z\sim2$. Either black holes and galaxies grow almost co-eval since $z\sim2$, or galaxies grow faster and the black holes would have to catch up to end up at the local relation at $z=0$. The latter is also suggested by observations of $z\sim2$ SMGs  \citep{Borys:2005,Alexander:2008,Carrera:2011}. Within $0<z<2$ most of the black hole growth is governed by secular processes and not by major mergers \citep{Cisternas:2011b,Schawinski:2011,Kocevski:2012}. These processes probably grow the black hole at the same rate as the spheroid component.
This changes at $z>2$, beyond the peak of AGN activity, where major mergers likely become the dominating black hole growth mechanism \citep{Draper:2012}. Therefore also a change in the evolution of the $M_\bullet-M_\ast$ relation could be expected at $z>2$. Currently, this redshift regime is still poorly covered by observations, and usually restricted to the most luminous, and hence most biased, QSOs. We illustrated the poor constraints that can be drawn from current observations, employing a $z\sim6$ QSO sample based on \citet{Wang:2010}. While the sample shows a large \textit{apparent} positive offset from the local relation, we find here that there is no \textit{intrinsic} offset, i.e. it is fully consistent with no or even negative evolution when sample selection is taken into account. We therefore conclude that even at $z>2$ and in particular at $z\sim6$ we currently have no statistically significant evidence for evolution in the  $M_\bullet-M_\ast$ relation. We confirmed these results by additionally studying the low redshift sample from \citet{Cisternas:2011} and the small $z\sim4$ sample from \citet{Targett:2012}. 

Future studies will reduce the observational uncertainties and increase the sample size and will allow a better assessment of the $M_\bullet-M_\ast$ relation at these early cosmic epochs. We will discuss the requirements on the design of these future studies of the $M_\bullet-M_\ast$ relation in this redshift regime in a companion paper.

While selection effects severely complicate the observational study of the  $M_\bullet-$bulge relation at high redshift, these can be included in the analysis as we have demonstrated here. We anticipate that the method presented here will contribute to an improved understanding of black hole - galaxy co-evolution.
However, essential prerequisites are (1) the use of a well defined sample, i.e. the selection function for the sample has to be under control, and (2) knowledge of the underlying demographics of galaxies and active black holes over the black hole mass and galaxy property range covered by the observations.
Given the expected future increase in the size of well defined samples and a better understanding of the underlying black hole and galaxy demographics, we anticipate that we will be able to trace back the cosmic history of the $M_\bullet-$bulge relations in the future.

\section*{Acknowledgements}
We thank Knud Jahnke for fruitful discussions on this work. We also thank the anonymous referee for an efficient and constructive report that strengthened our conclusions.
We acknowledge support by the Deutsche Forschungsgemeinschaft under its priority programme SPP1177, grant Wi~1369/23. 
AS  acknowledges support by the China Postdoctoral Scientific Foundation, grant 2012M510256.

\appendix

\section[]{Accounting for Upper limits} \label{sec:upperlimits}
For AGN samples, per definition the black hole is always detected and its mass is estimated via the virial method. However, the host galaxy is not always detected, as it can be outshined by the AGN. But, if at least upper limits on the host galaxy property can be set, these objects can be included in the fitting. Otherwise their exclusion needs to be reflected in the selection function. 
We include upper limits on $x$ in the following manner, similar to \citet{Gultekin:2009}\footnote{We assume that every AGN is harboured by a galaxy, i.e. there are no ``naked`` quasars (See e.g. \citet{Magain:2005,Jahnke:2009b} on this issue)}.

First, we compute the probability $U(s_\mathrm{u},\mu)$ that a black hole with mass $\mu$ is located in a galaxy with their respective property greater than $s_\mathrm{u}$.
For the forward regression without measurement errors we simply assume a uniform distribution in $s$,
\begin{equation}
p(\mu \, |\, s_\mathrm{u},\er,z) = \frac{\int_{s_\mathrm{u}}^\infty \int \Omega\, p_\mathrm{ac}(\mu,z)\,g(\mu \, |\, s,z)\, \dd s}{\int \Omega\, p_\mathrm{ac}(\mu,z)\,g(\mu \, |\, s,z)\,\dd \mu} \ .
\end{equation}
For the inverse regression without measurement errors we get
\begin{equation}
 p(s_\mathrm{u} \, |\, \mu,z)  = \frac{\int_{s_\mathrm{u}}^\infty g(\mu \, |\, s)\, \Phi_s(s)\,\dd s }{\int g(\mu \, |\, s)\, \Phi_s(s) \dd s} \ . 
\end{equation} 
When we include measurement errors in $s$ and $\mu$,  the probability distribution for the forward regression is
\begin{equation}
 p(\mu_\mathrm{o} \, |\, s_\mathrm{o,u},\er,z)  =  
\frac{\int_{s_\mathrm{u}}^\infty \int \Omega\, p_\mathrm{ac}(\mu,z)\,g(\mu \, |\, s,z)\,\Phi_s(s)\,g(\mu_\mathrm{o} \, |\, \mu)\,\dd s \dd \mu}{\iint \Omega\, p_\mathrm{ac}(\mu,z)\,g(\mu \, |\, s,z)\,\Phi_s(s)\,\dd s \dd \mu} \ .
\end{equation} 
The likelihood for an upper limit at $1\sigma$ confidence level is 
\begin{equation}
 l_i = \delta_1 U(s_\mathrm{u},\mu) + (1-\delta_1)(1-U(s_\mathrm{u},\mu)) \ ,
\end{equation} 
where $U(s_\mathrm{u},\mu)$ is one of the above conditional probabilities and $\delta_1=0.16$ is the probability to find the galaxy having $s$ greater than $s_\mathrm{u}$. For an upper limit at a different confidence level this probability changes accordingly.

\bsp

\label{lastpage}


\begin{thebibliography}{}

\bibitem[Aird et al.(2012)]{Aird:2012} Aird, J., Coil, A.~L., 
Moustakas, J., et al.\ 2012, \apj, 746, 90 

\bibitem[Alexander et al.(2008)]{Alexander:2008} Alexander, D.~M., 
Brandt, W.~N., Smail, I., et al.\ 2008, \aj, 135, 1968

\bibitem[Angl{\'e}s-Alc{\'a}zar et al.(2013)]{Angles:2013} 
Angl{\'e}s-Alc{\'a}zar, D., {\"O}zel, F., 
\& Dav{\'e}, R.\ 2013, \apj, 770, 5 


\bibitem[{{Bell} {et~al.}(2003){Bell}, {McIntosh}, {Katz}, \&  {Weinberg}}]{Bell:2003}{Bell}, E.~F., {McIntosh}, D.~H., {Katz}, N., \& {Weinberg}, M.~D. 2003, \apjs, 149, 289

\bibitem[{{Bennert} {et~al.}(2010){Bennert}, {Treu}, {Woo}, {Malkan}, {Le
  Bris}, {Auger}, {Gallagher}, \& {Blandford}}]{Bennert:2010}
{Bennert}, V.~N., {Treu}, T., {Woo}, J., {et~al.} 2010, \apj, 708, 1507

\bibitem[Bezanson et al.(2011)]{Bezanson:2011} Bezanson, R., van 
Dokkum, P.~G., Franx, M., et al.\ 2011, \apjl, 737, L31

\bibitem[Bongiorno et al.(2012)]{Bongiorno:2012} Bongiorno, A., Merloni, A., Brusa, M., et al.\ 2012, \mnras, 427, 3103 

\bibitem[{{Booth} \& {Schaye}(2011)}]{Booth:2010}
{Booth}, C.~M. \& {Schaye}, J. 2011, \mnras, 413, 1158

\bibitem[Borys et al.(2005)]{Borys:2005} Borys, C., Smail, I., 
Chapman, S.~C., et al.\ 2005, \apj, 635, 853 

\bibitem[Canalizo et al.(2012)]{Canalizo:2012} Canalizo, G., Wold, 
M., Hiner, K.~D., et al.\ 2012, \apj, 760, 38

\bibitem[Carrera et al.(2011)]{Carrera:2011} Carrera, F.~J., Page, 
M.~J., Stevens, J.~A., et al.\ 2011, \mnras, 413, 2791 

\bibitem[Chae(2010)]{Chae:2010} Chae, K.-H.\ 2010, \mnras, 402, 
2031 

\bibitem[Cisternas et al.(2011)]{Cisternas:2011} Cisternas, M., Jahnke, K., Bongiorno, A., et al.\ 2011, \apjl, 741, L11 

\bibitem[Cisternas et al.(2011b)]{Cisternas:2011b} Cisternas, M., 
Jahnke, K., Inskip, K.~J., et al.\ 2011, \apj, 726, 57 


\bibitem[{{Croton}(2006)}]{Croton:2006}
{Croton}, D.~J. 2006, \mnras, 369, 1808

\bibitem[Daddi et al.(2005)]{Daddi:2005} Daddi, E., Renzini, A., 
Pirzkal, N., et al.\ 2005, \apj, 626, 680 


\bibitem[{{Decarli} {et~al.}(2010){Decarli}, {Falomo}, {Treves}, {Labita},
  {Kotilainen}, \& {Scarpa}}]{Decarli:2010}
{Decarli}, R., {Falomo}, R., {Treves}, A., {et~al.} 2010, \mnras, 402, 2453

\bibitem[{{Di Matteo} {et~al.}(2008){Di Matteo}, {Colberg}, {Springel},
  {Hernquist}, \& {Sijacki}}]{DiMatteo:2008}
{Di Matteo}, T., {Colberg}, J., {Springel}, V., {Hernquist}, L., \& {Sijacki},D. 2008, \apj, 676, 33


\bibitem[{{Di Matteo} {et~al.}(2005){Di Matteo}, {Springel}, \&
  {Hernquist}}]{DiMatteo:2005}
{Di Matteo}, T., {Springel}, V., \& {Hernquist}, L. 2005, \nat, 433, 604

\bibitem[Draper 
\& Ballantyne(2012)]{Draper:2012} Draper, A.~R., \& Ballantyne, D.~R.\ 2012, \apj, 751, 72

\bibitem[{{Dubois} {et~al.}(2011){Dubois}, {Devriendt}, {Slyz}, \&
  {Teyssier}}]{Dubois:2011}Dubois, Y., Devriendt, J., Slyz, A., \& Teyssier, R.\ 2012, \mnras, 420, 2662

\bibitem[{{Fan} {et~al.}(2001){Fan}, {Narayanan}, {Lupton}, {Strauss}, {Knapp},
  {Becker}, {White}, {Pentericci}, {Leggett}, {Haiman}, {Gunn}, {Ivezi{\'c}},
  {Schneider}, {Anderson}, {Brinkmann}, {Bahcall}, {Connolly}, {Csabai}, {Doi},
  {Fukugita}, {Geballe}, {Grebel}, {Harbeck}, {Hennessy}, {Lamb}, {Miknaitis},
  {Munn}, {Nichol}, {Okamura}, {Pier}, {Prada}, {Richards}, {Szalay}, \&
  {York}}]{Fan:2001}
{Fan}, X., {Narayanan}, V.~K., {Lupton}, R.~H., {et~al.} 2001, \aj, 122, 2833

\bibitem[{{Fan} {et~al.}(2006){Fan}, {Strauss}, {Richards}, {Hennawi},
  {Becker}, {White}, {Diamond-Stanic}, {Donley}, {Jiang}, {Kim}, {Vestergaard},
  {Young}, {Gunn}, {Lupton}, {Knapp}, {Schneider}, {Brandt}, {Bahcall},
  {Barentine}, {Brinkmann}, {Brewington}, {Fukugita}, {Harvanek}, {Kleinman},
  {Krzesinski}, {Long}, {Neilsen}, {Nitta}, {Snedden}, \& {Voges}}]{Fan:2006}
{Fan}, X., {Strauss}, M.~A., {Richards}, G.~T., {et~al.} 2006, \aj, 131, 1203

\bibitem[{{Fan} {et~al.}(2003){Fan}, {Strauss}, {Schneider}, {Becker}, {White},
  {Haiman}, {Gregg}, {Pentericci}, {Grebel}, {Narayanan}, {Loh}, {Richards},
  {Gunn}, {Lupton}, {Knapp}, {Ivezi{\'c}}, {Brandt}, {Collinge}, {Hao},
  {Harbeck}, {Prada}, {Schaye}, {Strateva}, {Zakamska}, {Anderson},
  {Brinkmann}, {Bahcall}, {Lamb}, {Okamura}, {Szalay}, \& {York}}]{Fan:2003}
{Fan}, X., {Strauss}, M.~A., {Schneider}, D.~P., {et~al.} 2003, \aj, 125, 1649

\bibitem[{{Fan} {et~al.}(2000){Fan}, {White}, {Davis}, {Becker}, {Strauss},
  {Haiman}, {Schneider}, {Gregg}, {Gunn}, {Knapp}, {Lupton}, {Anderson},
  {Anderson}, {Annis}, {Bahcall}, {Boroski}, {Brunner}, {Chen}, {Connolly},
  {Csabai}, {Doi}, {Fukugita}, {Hennessy}, {Hindsley}, {Ichikawa},
  {Ivezi{\'c}}, {Loveday}, {Meiksin}, {McKay}, {Munn}, {Newberg}, {Nichol},
  {Okamura}, {Pier}, {Sekiguchi}, {Shimasaku}, {Stoughton}, {Szalay},
  {Szokoly}, {Thakar}, {Vogeley}, \& {York}}]{Fan:2000}
{Fan}, X., {White}, R.~L., {Davis}, M., {et~al.} 2000, \aj, 120, 1167

\bibitem[{{Ferrarese} \& {Merritt}(2000)}]{Ferrarese:2000}
{Ferrarese}, L. \& {Merritt}, D. 2000, \apjl, 539, L9

\bibitem[{{Fontana} {et~al.}(2006){Fontana}, {Salimbeni}, {Grazian},
  {Giallongo}, {Pentericci}, {Nonino}, {Fontanot}, {Menci}, {Monaco},
  {Cristiani}, {Vanzella}, {de Santis}, \& {Gallozzi}}]{Fontana:2006}
{Fontana}, A., {Salimbeni}, S., {Grazian}, A., {et~al.} 2006, \aap, 459, 745

\bibitem[{{Gebhardt} {et~al.}(2000){Gebhardt}, {Bender}, {Bower}, {Dressler},
  {Faber}, {Filippenko}, {Green}, {Grillmair}, {Ho}, {Kormendy}, {Lauer},
  {Magorrian}, {Pinkney}, {Richstone}, \& {Tremaine}}]{Gebhardt:2000}
{Gebhardt}, K., {Bender}, R., {Bower}, G., {et~al.} 2000, \apjl, 539, L13

\bibitem[Gebhardt et al.(2000b)]{Gebhardt:2000b} Gebhardt, K., Kormendy, J., Ho, L.~C., et al.\ 2000, \apjl, 543, L5 

\bibitem[Graham et al.(2011)]{Graham:2011} Graham, A.~W., Onken,  C.~A., Athanassoula, E., \& Combes, F.\ 2011, \mnras, 412, 2211

\bibitem[{{G{\"u}ltekin} {et~al.}(2009){G{\"u}ltekin}, {Richstone}, {Gebhardt},
  {Lauer}, {Tremaine}, {Aller}, {Bender}, {Dressler}, {Faber}, {Filippenko},
  {Green}, {Ho}, {Kormendy}, {Magorrian}, {Pinkney}, \&
  {Siopis}}]{Gultekin:2009}
{G{\"u}ltekin}, K., {Richstone}, D.~O., {Gebhardt}, K., {et~al.} 2009, \apj,
  698, 198

\bibitem[{{H{\"a}ring} \& {Rix}(2004)}]{Haering:2004}
{H{\"a}ring}, N. \& {Rix}, H.-W. 2004, \apjl, 604, L89


\bibitem[Hiner et al.(2012)]{Hiner:2012} Hiner, K.~D., Canalizo, 
G., Wold, M., Brotherton, M.~S., \& Cales, S.~L.\ 2012, \apj, 756, 162

\bibitem[{{Ho}(2007)}]{Ho:2007}
{Ho}, L.~C. 2007, \apj, 669, 821

\bibitem[{{Hopkins} {et~al.}(2006){Hopkins}, {Robertson}, {Krause},
  {Hernquist}, \& {Cox}}]{Hopkins:2006}
{Hopkins}, P.~F., {Robertson}, B., {Krause}, E., {Hernquist}, L., \& {Cox},
  T.~J. 2006, \apj, 652, 107

\bibitem[{{Jahnke} {et~al.}(2009{\natexlab{a}}){Jahnke}, {Bongiorno}, {Brusa},
  {Capak}, {Cappelluti}, {Cisternas}, {Civano}, {Colbert}, {Comastri}, {Elvis},
  {Hasinger}, {Ilbert}, {Impey}, {Inskip}, {Koekemoer}, {Lilly}, {Maier},
  {Merloni}, {Riechers}, {Salvato}, {Schinnerer}, {Scoville}, {Silverman},
  {Taniguchi}, {Trump}, \& {Yan}}]{Jahnke:2009}
{Jahnke}, K., {Bongiorno}, A., {Brusa}, M., {et~al.} 2009{\natexlab{a}}, \apjl,
  706, L215

\bibitem[{{Jahnke} {et~al.}(2009{\natexlab{b}}){Jahnke}, {Elbaz}, {Pantin},
  {B{\"o}hm}, {Wisotzki}, {Letawe}, {Chantry}, \& {Lagage}}]{Jahnke:2009b}
{Jahnke}, K., {Elbaz}, D., {Pantin}, E., {et~al.} 2009{\natexlab{b}}, \apj,
  700, 1820

\bibitem[{{Jahnke} \& {Macci{\`o}}(2011)}]{Jahnke:2010}
{Jahnke}, K. \& {Macci{\`o}}, A.~V. 2011, \apj, 734, 92

\bibitem[{{Kauffmann} \& {Haehnelt}(2000)}]{Kauffmann:2000}
{Kauffmann}, G. \& {Haehnelt}, M. 2000, \mnras, 311, 576

\bibitem[Kelly \& Shen(2013)]{Kelly:2013} Kelly, B.~C., \& Shen, Y.\ 2013, \apj, 764, 45

\bibitem[Kocevski et al.(2012)]{Kocevski:2012} Kocevski, D.~D., 
Faber, S.~M., Mozena, M., et al.\ 2012, \apj, 744, 148

\bibitem[Kormendy \& Ho(2013)]{Kormendy:2013} Kormendy, J., \& Ho, L.~C.\ 2013, arXiv:1304.7762

\bibitem[{{Lamastra} {et~al.}(2010){Lamastra}, {Menci}, {Maiolino}, {Fiore}, \& {Merloni}}]{Lamastra:2010}{Lamastra}, A., {Menci}, N., {Maiolino}, R., {Fiore}, F., \& {Merloni}, A.  2010, \mnras, 405, 29

\bibitem[{{Lauer} {et~al.}(2007){Lauer}, {Tremaine}, {Richstone}, \&
  {Faber}}]{Lauer:2007}
{Lauer}, T.~R., {Tremaine}, S., {Richstone}, D., \& {Faber}, S.~M. 2007, \apj,
  670, 249

\bibitem[Lee et al.(2012)]{Lee:2012} Lee, K.-S., Ferguson, H.~C., Wiklind, T., et al.\ 2012, \apj, 752, 66


\bibitem[{{Magain} {et~al.}(2005){Magain}, {Letawe}, {Courbin}, {Jablonka},
  {Jahnke}, {Meylan}, \& {Wisotzki}}]{Magain:2005}
{Magain}, P., {Letawe}, G., {Courbin}, F., {et~al.} 2005, \nat, 437, 381

\bibitem[{{Magorrian} {et~al.}(1998){Magorrian}, {Tremaine}, {Richstone},
  {Bender}, {Bower}, {Dressler}, {Faber}, {Gebhardt}, {Green}, {Grillmair},
  {Kormendy}, \& {Lauer}}]{Magorrian:1998}
{Magorrian}, J., {Tremaine}, S., {Richstone}, D., {et~al.} 1998, \aj, 115, 2285

\bibitem[Maiolino et al.(2007)]{Maiolino:2007} Maiolino, R., Neri, R., Beelen, A., et al.\ 2007, \aap, 472, L33

\bibitem[{{Marconi} \& {Hunt}(2003)}]{Marconi:2003}
{Marconi}, A. \& {Hunt}, L.~K. 2003, \apjl, 589, L21

\bibitem[{{Marconi} {et~al.}(2004){Marconi}, {Risaliti}, {Gilli}, {Hunt},
  {Maiolino}, \& {Salvati}}]{Marconi:2004}
{Marconi}, A., {Risaliti}, G., {Gilli}, R., {et~al.} 2004, \mnras, 351, 169

\bibitem[McConnell \& Ma(2013)]{McConnell:2013} McConnell, N.~J., \& Ma, C.-P.\ 2013, \apj, 764, 184

\bibitem[{{McLure} \& {Jarvis}(2002)}]{McLure:2002}
{McLure}, R.~J. \& {Jarvis}, M.~J. 2002, \mnras, 337, 109

\bibitem[{{McLure} {et~al.}(2006){McLure}, {Jarvis}, {Targett}, {Dunlop}, \&
  {Best}}]{McLure:2006}
{McLure}, R.~J., {Jarvis}, M.~J., {Targett}, T.~A., {Dunlop}, J.~S., \& {Best},
  P.~N. 2006, \mnras, 368, 1395

\bibitem[{{Merloni} {et~al.}(2004){Merloni}, {Rudnick}, \& {Di
  Matteo}}]{Merloni:2004}
{Merloni}, A., {Rudnick}, G., \& {Di Matteo}, T. 2004, \mnras, 354, L37

\bibitem[{{Merloni} {et~al.}(2010){Merloni}, {Bongiorno}, {Bolzonella},
  {Brusa}, {Civano}, {Comastri}, {Elvis}, {Fiore}, {Gilli}, {Hao}, {Jahnke},
  {Koekemoer}, {Lusso}, {Mainieri}, {Mignoli}, {Miyaji}, {Renzini}, {Salvato},
  {Silverman}, {Trump}, {Vignali}, {Zamorani}, {Capak}, {Lilly}, {Sanders},
  {Taniguchi}, {Bardelli}, {Carollo}, {Caputi}, {Contini}, {Coppa}, {Cucciati},
  {de la Torre}, {de Ravel}, {Franzetti}, {Garilli}, {Hasinger}, {Impey},
  {Iovino}, {Iwasawa}, {Kampczyk}, {Kneib}, {Knobel}, {Kova{\v c}},
  {Lamareille}, {Le Borgne}, {Le Brun}, {Le F{\`e}vre}, {Maier}, {Pello},
  {Peng}, {Perez Montero}, {Ricciardelli}, {Scodeggio}, {Tanaka}, {Tasca},
  {Tresse}, {Vergani}, \& {Zucca}}]{Merloni:2010}
{Merloni}, A., {Bongiorno}, A., {Bolzonella}, M., {et~al.} 2010, \apj, 708, 137

\bibitem[{{Merloni} \& {Heinz}(2008)}]{Merloni:2008} {Merloni}, A. \& {Heinz}, S. 2008, \mnras, 388, 1011

\bibitem[{{Nesvadba} {et~al.}(2011){Nesvadba}, {De Breuck}, {Lehnert}, {Best},
  {Binette}, \& {Proga}}]{Nesvadba:2011}
{Nesvadba}, N.~P.~H., {De Breuck}, C., {Lehnert}, M.~D., {et~al.} 2011, \aap,
  525, A43+
  
\bibitem[Nobuta et al.(2012)]{Nobuta:2012} Nobuta, K., Akiyama, M., 
Ueda, Y., et al.\ 2012, \apj, 761, 143

\bibitem[Park et al.(2012)]{Park:2012} Park, D., Kelly, B.~C.,  Woo, J.-H., \& Treu, T.\  2012, \apjs, 203, 6 

\bibitem[{{Peng}(2007)}]{Peng:2007}
{Peng}, C.~Y. 2007, \apj, 671, 1098

\bibitem[{{Peng} {et~al.}(2006){Peng}, {Impey}, {Rix}, {Kochanek}, {Keeton},
  {Falco}, {Leh{\'a}r}, \& {McLeod}}]{Peng:2006b}
{Peng}, C.~Y., {Impey}, C.~D., {Rix}, H., {et~al.} 2006, \apj, 649, 616

\bibitem[Portinari et al.(2012)]{Portinari :2012} Portinari, L., Kotilainen, J., Falomo, R., \& Decarli, R.\ 2012, \mnras, 420, 732 

\bibitem[Richards et al.(2006a)]{Richards:2006a} Richards, G.~T., Lacy, 
M., Storrie-Lombardi, L.~J., et al.\ 2006, \apjs, 166, 470

\bibitem[Richards et al.(2006b)]{Richards:2006b} Richards, G.~T., 
Strauss, M.~A., Fan, X., et al.\ 2006, \aj, 131, 2766 

\bibitem[{{Riechers} {et~al.}(2008){Riechers}, {Walter}, {Brewer}, {Carilli},
  {Lewis}, {Bertoldi}, \& {Cox}}]{Riechers:2008}
{Riechers}, D.~A., {Walter}, F., {Brewer}, B.~J., {et~al.} 2008, \apj, 686, 851

\bibitem[{{Robertson} {et~al.}(2006){Robertson}, {Hernquist}, {Cox}, {Di
  Matteo}, {Hopkins}, {Martini}, \& {Springel}}]{Robertson:2006}
{Robertson}, B., {Hernquist}, L., {Cox}, T.~J., {et~al.} 2006, \apj, 641, 90

\bibitem[{{Salviander} {et~al.}(2007){Salviander}, {Shields}, {Gebhardt}, \&
  {Bonning}}]{Salviander:2007}
{Salviander}, S., {Shields}, G.~A., {Gebhardt}, K., \& {Bonning}, E.~W. 2007,
  \apj, 662, 131
  
 \bibitem[Salviander \& Shields(2013)]{Salviander:2013} Salviander, S., \& Shields, G.~A.\ 2013, \apj, 764, 80
 
 
 \bibitem[Schawinski et al.(2011)]{Schawinski:2011} Schawinski, K., 
Treister, E., Urry, C.~M., et al.\ 2011, \apjl, 727, L31 

\bibitem[{{Schramm} {et~al.}(2008){Schramm}, {Wisotzki}, \&
  {Jahnke}}]{Schramm:2008}
{Schramm}, M., {Wisotzki}, L., \& {Jahnke}, K. 2008, \aap, 478, 311

\bibitem[Schramm  \& Silverman(2013)]{Schramm:2013} Schramm, M., \& Silverman, J.~D.\ 2013, \apj, 767, 13 

\bibitem[Schulze \& Gebhardt(2011)]{Schulze:2011} Schulze, A., \& Gebhardt, K. 2011, \apj, 729, 21

\bibitem[{{Schulze} \& {Wisotzki}(2010)}]{Schulze:2010} {Schulze}, A. \& {Wisotzki}, L. 2010, \aap, 516, A87+

\bibitem[{{Schulze} \& {Wisotzki}(2011)}]{Schulze:2011b}
{Schulze}, A. \& {Wisotzki}, L. 2011, \aap, 535, A87

\bibitem[{{Shankar} {et~al.}(2009){Shankar}, {Bernardi}, \&
  {Haiman}}]{Shankar:2009}
{Shankar}, F., {Bernardi}, M., \& {Haiman}, Z. 2009, \apj, 694, 867

\bibitem[{{Shen} \& {Kelly}(2010)}]{Shen:2010}{Shen}, Y. \& {Kelly}, B.~C. 2010, \apj, 713, 41

\bibitem[{{Sheth} {et~al.}(2003){Sheth}, {Bernardi}, {Schechter}, {Burles},
  {Eisenstein}, {Finkbeiner}, {Frieman}, {Lupton}, {Schlegel}, {Subbarao},
  {Shimasaku}, {Bahcall}, {Brinkmann}, \& {Ivezi{\'c}}}]{Sheth:2003}
{Sheth}, R.~K., {Bernardi}, M., {Schechter}, P.~L., {et~al.} 2003, \apj, 594,
  225

\bibitem[{{Shields} {et~al.}(2003){Shields}, {Gebhardt}, {Salviander}, {Wills},
  {Xie}, {Brotherton}, {Yuan}, \& {Dietrich}}]{Shields:2003}
{Shields}, G.~A., {Gebhardt}, K., {Salviander}, S., {et~al.} 2003, \apj, 583,
  124
  
  \bibitem[{{Sijacki} {et~al.}(2007){Sijacki}, {Springel}, {Di Matteo}, \&
  {Hernquist}}]{Sijacki:2007}
{Sijacki}, D., {Springel}, V., {Di Matteo}, T., \& {Hernquist}, L. 2007,
  \mnras, 380, 877

\bibitem[{{Silk} \& {Rees}(1998)}]{Silk:1998}
{Silk}, J. \& {Rees}, M.~J. 1998, \aap, 331, L1

\bibitem[{{Somerville} {et~al.}(2008){Somerville}, {Hopkins}, {Cox},
  {Robertson}, \& {Hernquist}}]{Somerville:2008}
{Somerville}, R.~S., {Hopkins}, P.~F., {Cox}, T.~J., {Robertson}, B.~E., \&
  {Hernquist}, L. 2008, \mnras, 391, 481
  
  \bibitem[Targett et al.(2012)]{Targett:2012} Targett, T.~A., Dunlop, 
J.~S., \& McLure, R.~J.\ 2012, \mnras, 420, 3621 

\bibitem[{{Tremaine} {et~al.}(2002){Tremaine}, {Gebhardt}, {Bender}, {Bower},
  {Dressler}, {Faber}, {Filippenko}, {Green}, {Grillmair}, {Ho}, {Kormendy},
  {Lauer}, {Magorrian}, {Pinkney}, \& {Richstone}}]{Tremaine:2002}
{Tremaine}, S., {Gebhardt}, K., {Bender}, R., {et~al.} 2002, \apj, 574, 740

\bibitem[Trujillo et al.(2007)]{Trujillo:2007} Trujillo, I., 
Conselice, C.~J., Bundy, K., et al.\ 2007, \mnras, 382, 109

\bibitem[van Dokkum et al.(2008)]{vanDokkum:2008} van Dokkum, P.~G., 
Franx, M., Kriek, M., et al.\ 2008, \apjl, 677, L5 

\bibitem[{{Vestergaard} \& {Peterson}(2006)}]{Vestergaard:2006}
{Vestergaard}, M. \& {Peterson}, B.~M. 2006, \apj, 641, 689

\bibitem[Volonteri  \& Stark(2011)]{Volonteri:2011} Volonteri, M., \& Stark, D.~P.\ 2011, \mnras, 417, 2085

\bibitem[{{Walter} {et~al.}(2004){Walter}, {Carilli}, {Bertoldi}, {Menten},
  {Cox}, {Lo}, {Fan}, \& {Strauss}}]{Walter:2004}
{Walter}, F., {Carilli}, C., {Bertoldi}, F., {et~al.} 2004, \apjl, 615, L17

\bibitem[{{Wang} {et~al.}(2010){Wang}, {Carilli}, {Neri}, {Riechers}, {Wagg},
  {Walter}, {Bertoldi}, {Menten}, {Omont}, {Cox}, \& {Fan}}]{Wang:2010}
{Wang}, R., {Carilli}, C.~L., {Neri}, R., {et~al.} 2010, \apj, 714, 699

\bibitem[Wang et al.(2013)]{Wang:2013} Wang, R., Wagg, J., Carilli, C.~L., et al.\ 2013,  \apj, 773, 44


\bibitem[Willott et al.(2013)]{Willott:2013} Willott, C.~J., Omont, 
A., \& Bergeron, J.\ 2013, \apj, 770, 13

\bibitem[{{Woo} {et~al.}(2008){Woo}, {Treu}, {Malkan}, \&
  {Blandford}}]{Woo:2008}
{Woo}, J.-H., {Treu}, T., {Malkan}, M.~A., \& {Blandford}, R.~D. 2008, \apj,
  681, 925
  
 \bibitem[Woo et al.(2013)]{Woo:2013} Woo, J.-H., Schulze, A., Park, D., et al.\ 2013, \apj, 772, 49
  
\bibitem[Zhang et al.(2012)]{Zhang:2012} Zhang, X., Lu, Y., \& Yu, Q.\ 2012, \apj, 761, 5 
\end{thebibliography}
\end{document}